\documentclass[aps,prc,twocolumn,superscriptaddress,floatfix]{revtex4-1}

\usepackage{graphicx}
\usepackage{physics}
\usepackage{epstopdf}
\usepackage{xcolor}
\newcommand{\comment}[1]{}

\begin{document}


\title{Spectroscopy and excited-state $\boldsymbol{g}$~factors in weakly collective $\boldsymbol{^{111}}$Cd: confronting collective and microscopic models}


\author{B.J.~Coombes}
\author{A.E.~Stuchbery}
\affiliation{Department of Nuclear Physics, Research School of Physics and Engineering, The Australian National University, Canberra, ACT 2601, Australia}
\author{A.~Blazhev}
\affiliation{Institut f{\"u}r Kernphysik, Universit{\"a}t zu K{\"o}ln, Z{\"u}lpicher Stra{\ss}e 77, D-50937 K{\"o}ln, Germany}
\author{H.~Grawe}
\affiliation{GSI Helmholtzzentrum f{\"u}r Schwerionenforschung GmbH, D-64291 Darmstadt, Germany}
\author{M.W.~Reed}
\author{A.~Akber}
\author{J.T.H.~Dowie}
\author{M.S.M.~Gerathy}
\author{T.J. Gray}
\author{T.~Kib\'edi}
\author{A.J.~Mitchell}
\author{T.~Palazzo}
\affiliation{Department of Nuclear Physics, Research School of Physics and Engineering, The Australian National University, Canberra, ACT 2601, Australia}


\date{\today}

\begin{abstract}

\begin{description}

\item[Background] The even cadmium isotopes near the neutron midshell have long been considered among the best examples of vibrational nuclei. However, the vibrational nature of these nuclei has been questioned based on $E2$ transition rates that are not consistent with vibrational excitations. In the neighbouring odd-mass nuclei, the $g$~factors of the low-excitation collective states have been shown to be more consistent with a deformed rotational core than a vibrational core. Moving beyond the comparison of vibrational versus rotational models, recent advances in computational power have made shell-model calculations feasible for Cd isotopes. These calculations may give insights into the emergence and nature of collectivity in the Cd isotopes.

\item[Purpose] To investigate the nature of collective excitations in the $A\sim100$ region through experimental and theoretical studies of magnetic moments and electromagnetic transitions in $^{111}$Cd.

\item[Method] The spectroscopy of $^{111}$Cd has been studied following Coulomb excitation. Angular correlation measurements, transient-field $g$-factor measurements and lifetime measurements by the Doppler-broadened line shape method were performed. The structure of the nucleus was explored in relation to particle-vibration versus particle-rotor interpretations. Large-scale shell-model calculations were performed with the SR88MHJM Hamiltonian.

\item[Results] Excited-state $g$~factors have been measured, spin assignments examined and lifetimes determined. Attention was given to the reported $5/2^{+}$ \mbox{753-keV} and $3/2^{+}$ \mbox{755-keV} states. The $3/2^{+}$ \mbox{755-keV} level was not observed; evidence is presented that the reported $3/2^{+}$ state was a misidentification of the $5/2^{+}$ \mbox{753-keV} state.

\item[Conclusions] It is shown that the $g$~factors and level structure of $^{111}$Cd are not readily explained by the particle-vibration model. A particle-rotor approach has both successes and limitations. The shell-model approach successfully reproduces much of the known low-excitation structure in $^{111}$Cd.

\end{description}
\end{abstract}


\maketitle

\section{Introduction}

The even cadmium isotopes with $Z=48$ and mass numbers near $A=110$ have long been considered good examples of spherical vibrational nuclei \cite{bohr-mottelson}. The discovery of non-zero electric quadrupole moments for the $2^{+}$ excited states~\cite{Esat1976,Gillespie1977,Maynard1977}, consistent with weak deformations of $\epsilon_{2}\sim0.15$, was handled by introducing anharmonicities into the vibrational model~\cite{Bes1969,Kotila2003}. Later, the presence of intruder states, due to the 2-particle 4-hole proton excitations near the nominal two-phonon triplet, was included through configuration mixing~\cite{Heyde1982,Jolie1990,Deleze1993}. Recently, however, it has been found that this interpretation, based on an anharmonic vibrator model with strong intruder mixing, is not consistent with experimental electromagnetic transition rates~\cite{Bandyopadhyay2007,Garrett2007,Garrett2008,Batchelder2009,Garrett2012,Batchelder2012}. The confrontation between theory and experiment in the Cd isotopes has sparked a more general debate over the nature of nuclei that are not clearly deformed rotors~\cite{Garrett2010,Wood2012}. Other work has contended that the Cd isotopes are consistent with a vibrational interpretation, at least to within the accuracy of the theoretical models~\cite{Aprahamian1987,Coello2015}.

New insights into this problem are now possible from a microscopic perspective because computation power has reached the point where the Cd isotopes toward the neutron mid-shell have become accessible to large-scale shell-model calculations. Schmidt \textit{et al.}~\cite{Schmidt2017} have recently used such calculations to characterize the onset of quadrupole collectivity in $^{98-108}$Cd. Similar calculations describe the ground-state electromagnetic moments in the odd-mass $^{101-109}$Cd isotopes~\cite{Yordanov2018}.

The present work, which includes both experiment and theory, concerns the odd-$A$ nuclide $^{111}$Cd. This nucleus lies between $^{110}$Cd and $^{112}$Cd, both of which have been considered textbook examples of vibrational nuclei \cite{bohr-mottelson,Casten2000}. It follows an earlier study in which it was shown that for both $^{111}$Cd and $^{113}$Cd, the $g$~factors of the $1/2^{+}$ ground-states, and the strongly Coulomb-excited low-lying $3/2^{+}$ and $5/2^{+}$~levels are consistent with a particle-plus-rotor description whereas the signs of the $g$~factors of the $5/2^{+}$ states are incompatible with the particle-vibrational model~\cite{Stuchbery2016}. Contrasting with this observation, a $3/2^{+}$ state at 681~keV in $^{113}$Cd was excited with sufficient intensity to determine its $g$~factor, giving $g=+1.4(4)$~\cite{Stuchbery2016}. Such a large $g$~factor in a collective odd-$N$ nuclide is difficult to explain unless it is associated with the coupling of an odd-neutron in the $\nu s_{1/2}$ orbit to a spherical $2^{+}$ core excitation. The present experimental study sought to identify the corresponding state in $^{111}$Cd. Two candidates are the $5/2^{+}$~state at 753~keV and a reported $3/2^{+}$~state at 755~keV~\cite{Blachot2009}.

\begin{figure*}[ht!]
\includegraphics[width=2.05\columnwidth]{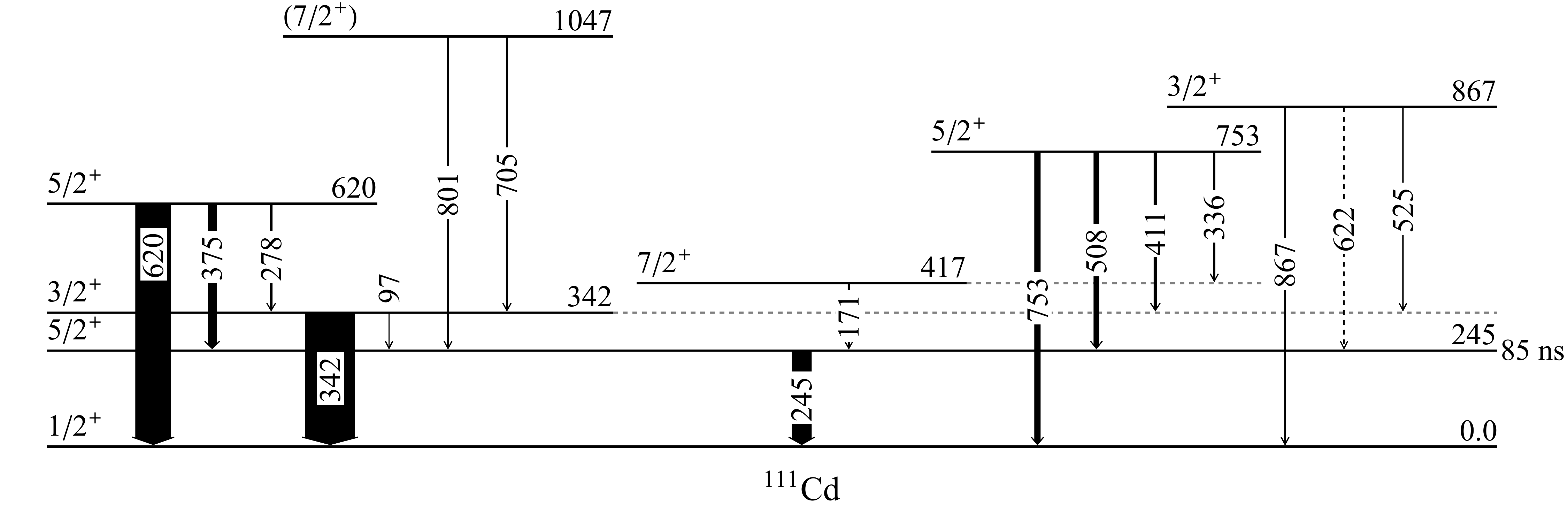}
\caption{Observed level scheme of $^{111}$Cd. Transition energies are given in keV and their intensities following Coulomb excitation with 90-MeV $^{32}$S beams are indicated by the arrow widths. The branching ratio for the dashed 622-keV line has been measured in Ref.~\cite{Krane2015} to be approximately twice that of the 867-keV transition. It was obscured in the present work by the much stronger 620-keV transition but could be resolved with poor statistics in particle-$\gamma$-$\gamma$~coincidences.}
\label{LevelScheme}
\end{figure*}

The present theoretical work evaluates both particle-vibration and particle-rotor interpretations of $^{111}$Cd. It also reports large-basis shell-model calculations which can begin to illuminate the nature of the collectivity in $^{111}$Cd from a microscopic perspective.

Following a description of the experiment in Sec.~\ref{ExperimentSection}, and a presentation of the analysis procedures and results in Sec.~\ref{AnalysisResultsSection}, the theoretical calculations are presented and discussed in Sec.~\ref{DiscussionSection}, followed by conclusions in Sec.~\ref{ConclusionSection}.

\section{Experiment} \label{ExperimentSection}

Experiments were performed using the Australian National University (ANU) Hyperfine Spectrometer~\cite{Stuchbery3000}. A 90-MeV $^{32}$S beam from the ANU 14UD Pelletron accelerator Coulomb excited nuclei in a 95\% enriched $^{111}$Cd target. Isotopic percentages are given in Table~\ref{IsotopicAssay}. A three-layer target was constructed. The first layer was  $\sim$0.36 mg/cm$^{2}$-thick enriched $^{111}$Cd electroplated onto the second 2.47 mg/cm$^{2}$ $^{\rm{nat}}$Fe layer. The third layer was a 12~$\mu$m thick natural copper layer. Evaporated indium ($\sim$0.3~mg/cm$^{2}$) was used to bond the Fe foil to the copper layer, which provided a field-free environment for Cd recoils and good thermal conduction. Cadmium has a low melting point of 321$^{\circ}$C. Therefore, in order to maintain structural integrity, the completed target was cooled by a Sumitomo RDK-408D cryocooler, which kept it at a constant temperature of $\sim$~6~K. The magnetic layer was polarized by a 0.09~T external magnetic field, the direction of which (`up' or `down') was reversed periodically, approximately every 15 minutes.

\begin{table}[b]
  \caption{Isotopic assay of the enriched $^{111}$Cd target. }
  \centering
  \begin{ruledtabular}
  \begin{tabular}{  c  c  c  c  c  c  c  c  c  }
  &  $^{106}$Cd  &  $^{108}$Cd  &  $^{110}$Cd  &  $^{111}$Cd  &  $^{112}$Cd  &  $^{113}$Cd  &  $^{114}$Cd  &  $^{116}$Cd \\
  \%  &  0.02  &  0.03  &  0.63  &  95.29  &  2.61  &  0.49  &  0.71  &  0.22 \\
  \end{tabular}
  \end{ruledtabular}
  \label{IsotopicAssay}
\end{table}

Two silicon photodiode detectors were placed above and below the beam axis, upstream from the target, to detect backscattered particles at average angles of $149^{\circ}$ to the beam direction. The vertical distance of these detectors from the beam axis was 4.6 mm, with 9.78 mm active height, 8.84 mm active width, and a horizontal distance back from the target of 16.2 mm.

Four high-purity germanium (HPGe) detectors were placed around the target chamber in the horizontal plane. The back-angle detectors were kept at $\pm115^{\circ}$ to the beam. The forward-detector angles varied to measure the particle-$\gamma$ angular correlations. Specifically, the forward $\gamma$-ray detectors were placed in turn at the angles $0^{\circ}$, $\pm32^{\circ}$, $\pm35^{\circ}$, $\pm45^{\circ}$, $\pm55^{\circ}$, $\pm65^{\circ}$, and $-85^{\circ}$ to the beam. The four $\gamma$-ray detectors were then fixed at $\theta_{\gamma}=\pm65^{\circ},\pm115^{\circ}$ for the $g$-factor measurements. These angles are near the maximum slope of the angular correlation for $E2$ multipolarity thus increasing the sensitivity of the $g$-factor measurement.

\section{Analysis and Results} \label{AnalysisResultsSection}

\subsection{$\gamma$-ray spectra and angular correlations}

\begin{figure*}[t!]
\includegraphics[height=1.5\columnwidth,angle=-90]{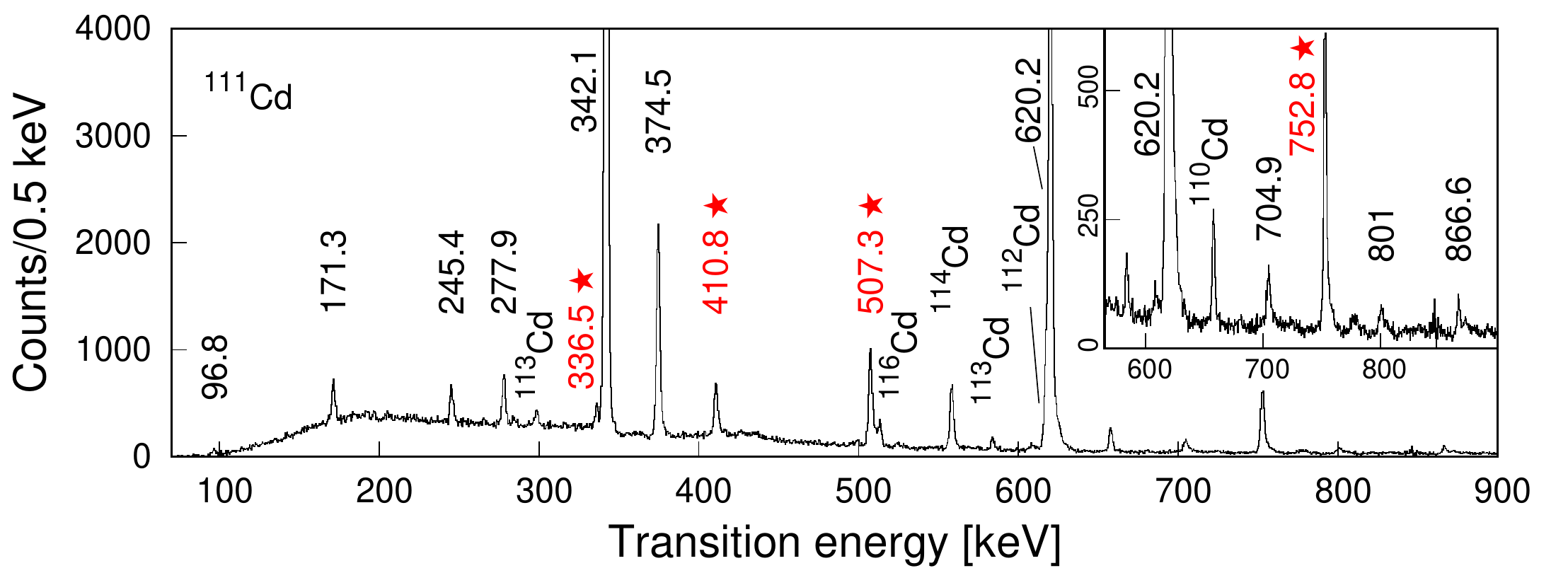}
\caption{Gamma-ray spectrum measured at $+65^{\circ}$ to the beam in coincidence with backscattered beam ions. All data taken at $+65^{\circ}$ during the transient field precession measurement are included. The inset expands the 565-900~keV region which shows a single strong peak at 752.8~keV and no peak at 754.9~keV. Transitions labeled in red and with a star are decays from the \mbox{752.8-keV} state.}
\label{gammaspectrum}
\end{figure*}

\begin{figure}[ht!]
\includegraphics[height=\columnwidth,angle=-90]{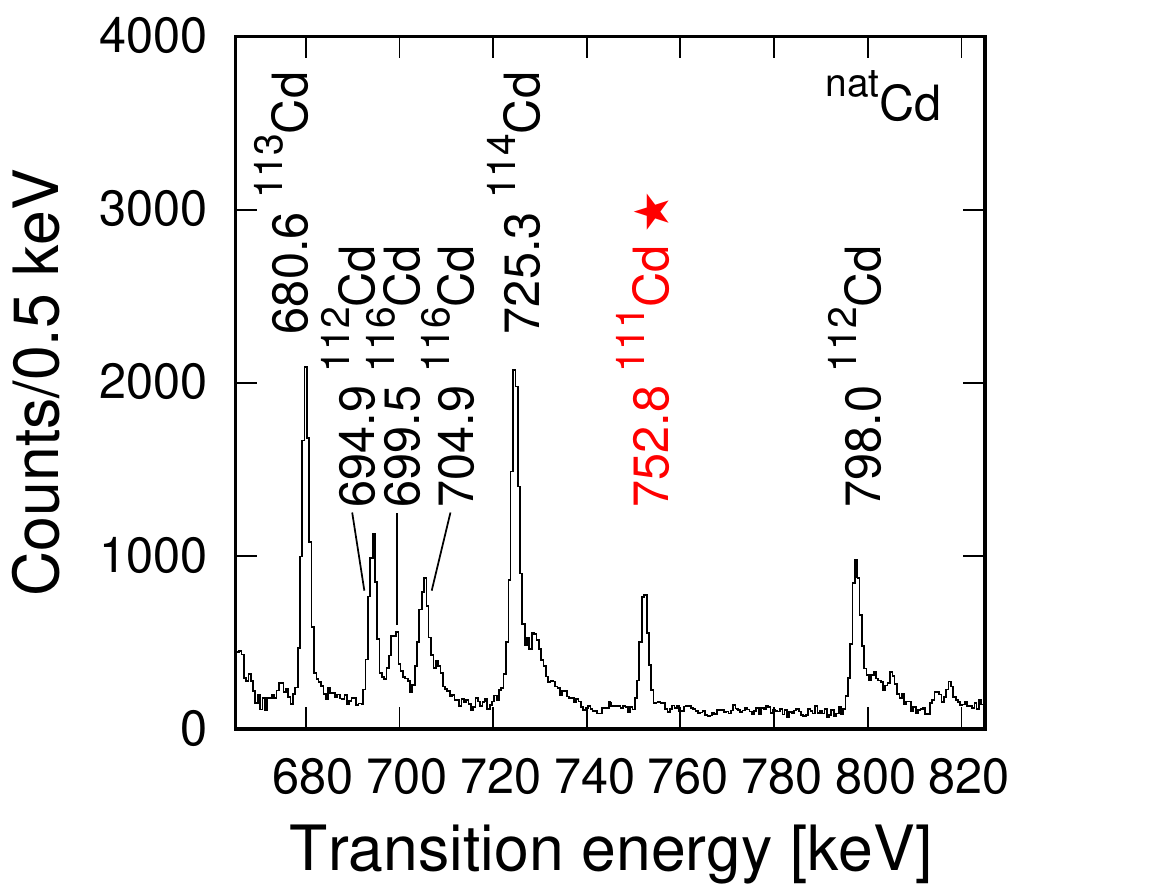}
\caption{Particle-$\gamma$ coincidence $\gamma$-ray spectrum at $+65^{\circ}$ to the beam for a natural Cd target from measurements reported in Ref.~\cite{Chamoli2011}. The red, starred transition corresponds to the single peak at 752.8~keV.}
\label{gammaspectrumChamoli}
\end{figure}

The observed level scheme is shown in Fig.~\ref{LevelScheme}. An example of the spectrum recorded by the $+65^{\circ}$ $\gamma$-ray detector in coincidence with backscattered beam ions is displayed in Fig.~\ref{gammaspectrum}. Each peak observed in the spectrum has been identified. A spectrum of transitions between 660 and 820~keV from a natural Cd target is shown in Fig.~\ref{gammaspectrumChamoli}. The transitions in $^{111}$Cd are listed in Table~\ref{Obstran} along with observed relative intensities and multipolarity assignments from the literature~\cite{Blachot2009}. Other peaks observed correspond to transitions in the neighboring even Cd isotopes and $^{113}$Cd present in the target (see Table~\ref{IsotopicAssay}).

\begin{figure}[ht]
\includegraphics[width=\columnwidth]{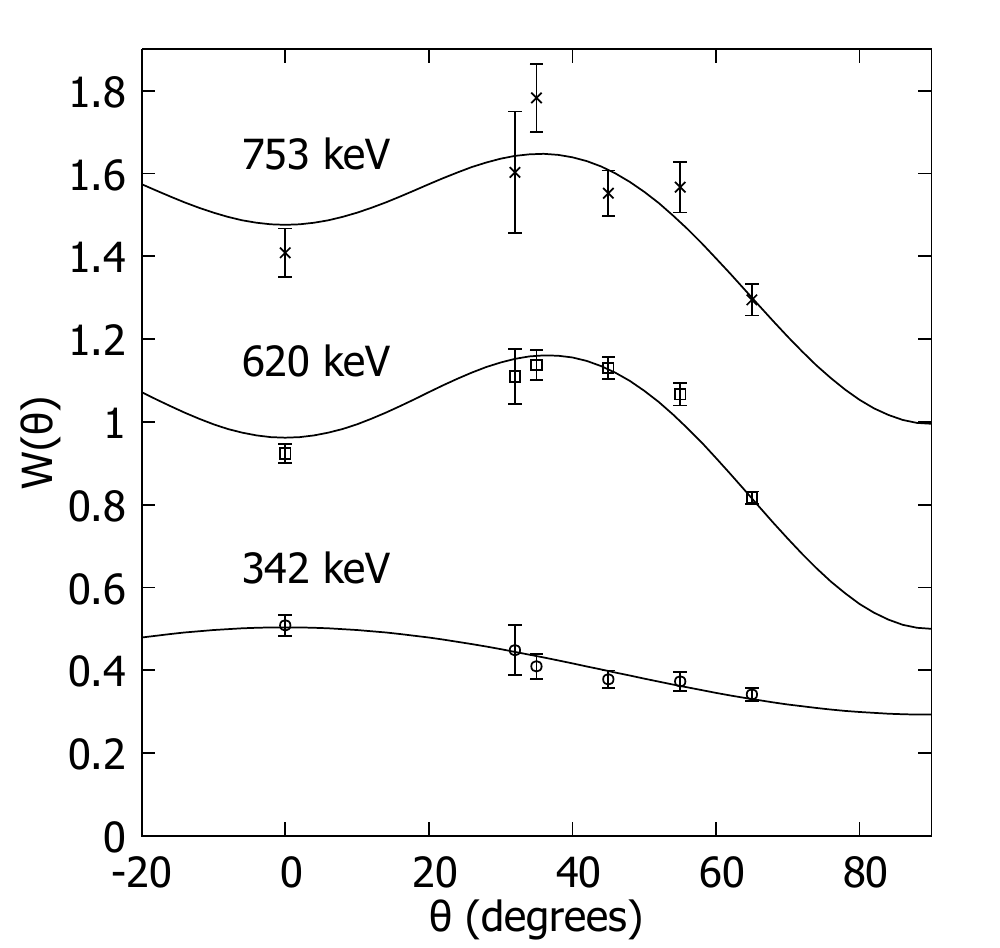}
\caption{Particle-$\gamma$ angular correlations for the \mbox{342-keV} (mixed $M1$/$E2$), \mbox{620-keV} ($E2$), and \mbox{753-keV} ($E2$) transitions, all of which are consistent with the multipolarities reported in the literature~\cite{Blachot2009}. Note the similar shape of the correlation for the 620- and \mbox{753-keV} $E2$ transitions and the contrast of those with the mixed multipolarity $M1$/$E2$ \mbox{342-keV} transition with mixing ratio $\delta=$+0.36~\cite{Blachot2009}. Correlations for the 342- and \mbox{753-keV} transitions have been vertically offset, respectively, by $-$0.6 and +0.5 for presentation.}
\label{angularcorrelations}
\end{figure}

Particle-$\gamma$ angular correlations were measured for the 342-, 620- and \mbox{753-keV} transitions. The counts in each peak at each angle were normalized to the sum of counts in the strong \mbox{342-keV} and \mbox{620-keV} transitions in the fixed ($+115^{\circ}$) detector. The angular correlations measured at positive angles are displayed in Fig.~\ref{angularcorrelations}. The negative-angle data points are consistent with the displayed positive-angle data.

\subsection{Excited states in $^{111}$Cd}

As clearly seen in Fig.~\ref{gammaspectrum}, only one peak is observed in the region near 755~keV. The energy of this peak agrees with that of the $5/2^{+}$ \mbox{752.8-keV} state listed in Nuclear Data Sheets~\cite{Blachot2009}. A \mbox{755-keV} transition was not observed. The angular correlation observed for the \mbox{753-keV} transition (Fig.~\ref{angularcorrelations}) implies $E2$ multipolarity and hence confirms the $5/2^{+}$ spin assignment for the \mbox{753-keV} state. It is noteworthy that the early Coulomb-excitation study of McDonald and Porter~\cite{McDonald1968} likewise reported a single peak near 750~keV. In their case the energy was given as 755~keV and the state was tentatively assigned a spin-parity of $3/2^{+}$ or $5/2^{+}$. Considering the results presented here, we conclude that the reported energy is in error by 2~keV, and that the same \mbox{753-keV} state was excited in Ref.~\cite{McDonald1968} as in the present experiment. Further evidence for this conclusion is presented in Fig.~\ref{gammaspectrumChamoli}, which shows the relevant part of the particle-$\gamma$ coincidence data following Coulomb excitation of a natural Cd target, taken from the work of Chamoli \textit{et al.}~\cite{Chamoli2011}. Displayed in Fig.~\ref{gammaspectrumChamoli} are many known transitions in Cd isotopes between 680 and 798~keV but, again, no evidence for a transition at 755~keV. It is worth noting that the observed branching ratios are in agreement with a recent decay measurement~\cite{Krane2015} which also observed a single transition at this energy.

\begin{ruledtabular}
\begin{table}[ht!]
  \caption{Observed level energies and spins, $\gamma$-ray transitions, and multipolarities taken from Nuclear Data Sheets~\cite{Blachot2009} along with intensities measured in the present work at $65^{\circ}$ to the beam, relative to the 342-keV transition.}
 \centering
  \def\arraystretch{1.5}
  \begin{tabular}{  c  c  c  c  c  c  c  }
    $E_{i}$   & $I^{+}_{i}$            & $E_{\gamma}$  & $E_{f}$  & $I^{+}_{f}$                     & $XL$   & $I_{\gamma}$ ($65^{\circ}$)   \\
     (keV)    &                    & (keV)         & (keV)    &                             &        &                \\ \hline
    0         & $\frac{1}{2}^{+}$  &               &  0       &                             &        &                \\
    245.4     & $\frac{5}{2}^{+}$  & 245.4         &  0       &     $\frac{1}{2}^{+}$       &  E2    & 387(40)        \\
    342.1     & $\frac{3}{2}^{+}$  & 342.1         &  0       &     $\frac{1}{2}^{+}$       & M1+E2  & 1000(18)       \\
              &                    & 97            &  245.4   &     $\frac{5}{2}^{+}$       & M1+E2  &                \\
    416.7     & $\frac{7}{2}^{+}$  & 171.3         &  245.4   &     $\frac{5}{2}^{+}$       & M1+E2  &  20(2)         \\
    620.18    & $\frac{5}{2}^{+}$  & 620.2         &  0       &     $\frac{1}{2}^{+}$       &  E2    & 715(30)        \\
              &                    & 374.7         &  245.4   &     $\frac{5}{2}^{+}$       & M1+E2  & 160(5)         \\
              &                    & 277.9         &  342.1   &     $\frac{3}{2}^{+}$       & M1+E2  & 30(2)          \\
    752.8     & $\frac{5}{2}^{+}$  & 752.8         &  0       &     $\frac{1}{2}^{+}$       &  E2    & 105(4)         \\
              &                    & 507.6         &  245.4   &     $\frac{5}{2}^{+}$       &        & 93(8)          \\
              &                    & 410.8         &  342.1   &     $\frac{3}{2}^{+}$       & M1+E2  & 46(3)          \\
              &                    & 336.5         &  416.7   &     $\frac{7}{2}^{+}$       &        & 16(3)          \\
    866.6     & $\frac{3}{2}^{+}$  & 866.6         &  0       &     $\frac{1}{2}^{+}$       & M1+E2  & 11(2)          \\
              &                    & 622           &  245.4   &     $\frac{5}{2}^{+}$       &(M1,E2) &                \\
              &                    & 524.2         &  342.1   &     $\frac{3}{2}^{+}$       & M1+E2  &  3(2)          \\
    1046.8    & ($\frac{7}{2}^{+}$)& 801           &  245.4   &     $\frac{5}{2}^{+}$       &        & 12(3)          \\
              &                    & 704.9         &  342.1   &     $\frac{3}{2}^{+}$       &        & 20(2)          \\
  \end{tabular}
  \label{Obstran}
\end{table}
\end{ruledtabular}

\subsection{Thin-foil transient-field $g$-factor measurements}\label{TranFieldSection}

When an ion with energies of several MeV traverses a polarized ferromagnet it experiences a strong transient hyperfine field of the order of a few kilotesla. This hyperfine field causes a precession of the nuclear spin at a frequency proportional to the $g$~factor, ${\omega_{L}=-g\frac{\mu_N}{\hbar} B_{\rm{TF}}}$, where $B_{\rm{TF}}$ is the transient magnetic field strength, which varies with the ion velocity as it slows in the ferromagnetic host. The net precession angle observed is $\Delta\theta=g\phi$, where
\begin{equation}\label{PhiDefinition}
\phi(\tau)=-\frac{\mu_{N}}{\hbar}\int_{t_{i}}^{t_{e}}B_{\rm{TF}}[v(t)]e^{-t/\tau}dt,
\end{equation}
where $t_{i}$ and $t_{e}$ are the times of entry into and exit from the ferromagnetic layer, respectively. Provided the nuclear mean life $\tau$ exceeds $t_{e}$, $\phi(\tau)$ is a weak function of $\tau$.
Experimentally, the precession angle is found from the changes in count rate for a pair of detectors placed at the angles $\pm\theta_{\gamma}$ with respect to the beam axis, as the direction of the magnetic field is reversed. Specifically,
\begin{equation}
\Delta\theta=\frac{\epsilon}{S},
\end{equation}
where $S$ is the logarithmic derivative of the angular correlation at the detection angle ($+\theta_{\gamma}$) and
\begin{equation}
\epsilon=\frac{1-\rho}{1+\rho},
\end{equation}
where,
\begin{equation}
\rho=\sqrt{\frac{N(\theta_{\gamma})\uparrow}{N(\theta_{\gamma})\downarrow}\times\frac{N(-\theta_{\gamma})\downarrow}{N(-\theta_{\gamma})\uparrow}}=\frac{W(\theta_{\gamma}-\Delta\theta)}{W(\theta_{\gamma}+\Delta\theta)}.
\end{equation}
In this equation $N(\theta_{\gamma})\uparrow(\downarrow)$ represents the counts in the detector at angle $+\theta_{\gamma}$ with field `up' (`down') and $W(\theta)$ is the amplitude of the particle-$\gamma$ angular correlation.

Table~\ref{Kinematics} summarizes the reaction kinematics relevant for the $g$-factor measurement. The transient magnetic field was evaluated using a parametrization proposed by Stuchbery \textit{et al.}~\cite{Stuchbery1980} for the Pd isotopes, which has since been shown to also be appropriate for Cd in iron~\cite{Chamoli2011}. As reported in Ref.~\cite{Chamoli2011}, the transient-field strength for Cd in iron is calibrated relative to $g(2^+_1;^{106}{\rm Pd}) = +0.393(23)$ and has an uncertainty of $\pm 6\%$. This uncertainty is small compared to the uncertainties in the precession measurements. The time-dependent velocity was found using stopping powers from Ziegler~\cite{Ziegler1980} and evaluated for ions moving through the ferromagnetic foil. The even Cd isotopes have independently determined $g$~factors, however they constitute too small a fraction of the composition of the separated isotope target to serve as a calibration of the transient-field strength in the present measurement. Experimental $g$-factor results are presented in Table~\ref{gfactable}, along with previous measurements. Where previous measurements exist~\cite{Benczer1989,Stuchbery2016} agreement is found within experimental uncertainties.

\begin{ruledtabular}
\begin{table*}[t!]
  \caption{Reaction kinematics for the transient-field $g$-factor measurement. $E$, $I^{\pi}$, and $\tau$ are the energy, the spin and parity, and the lifetime of the excited state. $\langle E_{i}\rangle$ and $\langle E_{e}\rangle$ are the average energy of recoiling target ions entering and leaving the ferromagnetic layer, respectively. The corresponding entry and exit average velocities are given by $\langle v_{i}/v_{0}\rangle$ and $\langle v_{e}/v_{0}\rangle$ in units of the Bohr velocity ($v_{0}=c/137$). $\langle v/v_{0}\rangle$ is the average velocity of the ion as it passes through the ferromagnetic layer. The effective time for target ions passing through the ferromagnetic layer accounting for in-flight decays of the state is $t_{\rm{Fe}}$. The parameter $\phi(\tau)$ is proportional to the integrated transient magnetic field strength as defined in Eq.~(\ref{PhiDefinition}); it has an uncertainty of $\pm 6\%$ (see text).}
 \centering
  \def\arraystretch{1.5}
  \begin{tabular}{  c  c  c  c  c  c  c  c  c  c  }
  $E$            & $I^{\pi}$         & $\tau$ &     $\langle E_{i}\rangle$  & $\langle E_{e}\rangle$  & $\langle v_{i}/v_{0}\rangle$  & $\langle v_{e}/v_{0}\rangle$ & $\langle v/v_{0}\rangle$ & $t_{\rm{Fe}}$ & $-\phi(\tau)$  \\
     (keV)       &                   &  (ps)  &       (MeV)                 &  (MeV)                  &                               &                              &                          & (fs)          & (mrad)        \\ \hline
    342.1        & $\frac{3}{2}^{+}$ &  24(3) &     55.47                   & 10.97                   & 4.49                          &  2.00                        &  3.01                    & 481           & 44.1         \\
    620.2       & $\frac{5}{2}^{+}$ &   9.6(18)  &     55.45                   & 10.96                   & 4.49                          &  1.99                        &  3.01                    & 476           & 43.6         \\
    752.8        & $\frac{5}{2}^{+}$ & 4.0(16) &     55.44                   & 10.96                   & 4.49                          &  1.99                        &  3.03                    & 453           & 41.6         \\
  \end{tabular}
  \label{Kinematics}
\end{table*}
\end{ruledtabular}

\subsection{DBLS lifetime measurements}

The lifetime of the \mbox{753-keV} state was determined by the Doppler-broadened line shape method \cite{annurev.ns.18.120168.001405,Fossan-Warburton,Alexander-Forster,INAMURA1975529}. A computer program based on that of Wells and Johnson~\cite{Wells1991} was developed for this analysis.  A description of the code will be published elsewhere~\cite{Coombes3000}. The measured lifetimes are presented in Table~\ref{MeasuredLifetimes}. Individual $^{111}$Cd recoil events were simulated by Monte Carlo methods, accounting for the particle-detector geometry, reaction kinematics and the multilayered target. The distribution of Doppler shifts as a function of time and $\gamma$-ray detector acceptance angles, taking into account the \mbox{particle-$\gamma$} angular correlation, was then used to weight the evaluated line shape. As the $\gamma$-ray detectors were placed near the maximum slope of the angular correlations, the $\gamma$-ray intensity can vary significantly across the face of the detector, which can significantly affect the Doppler-broadened line shape.

\begin{figure}
\centerline{
\includegraphics[width=\columnwidth]{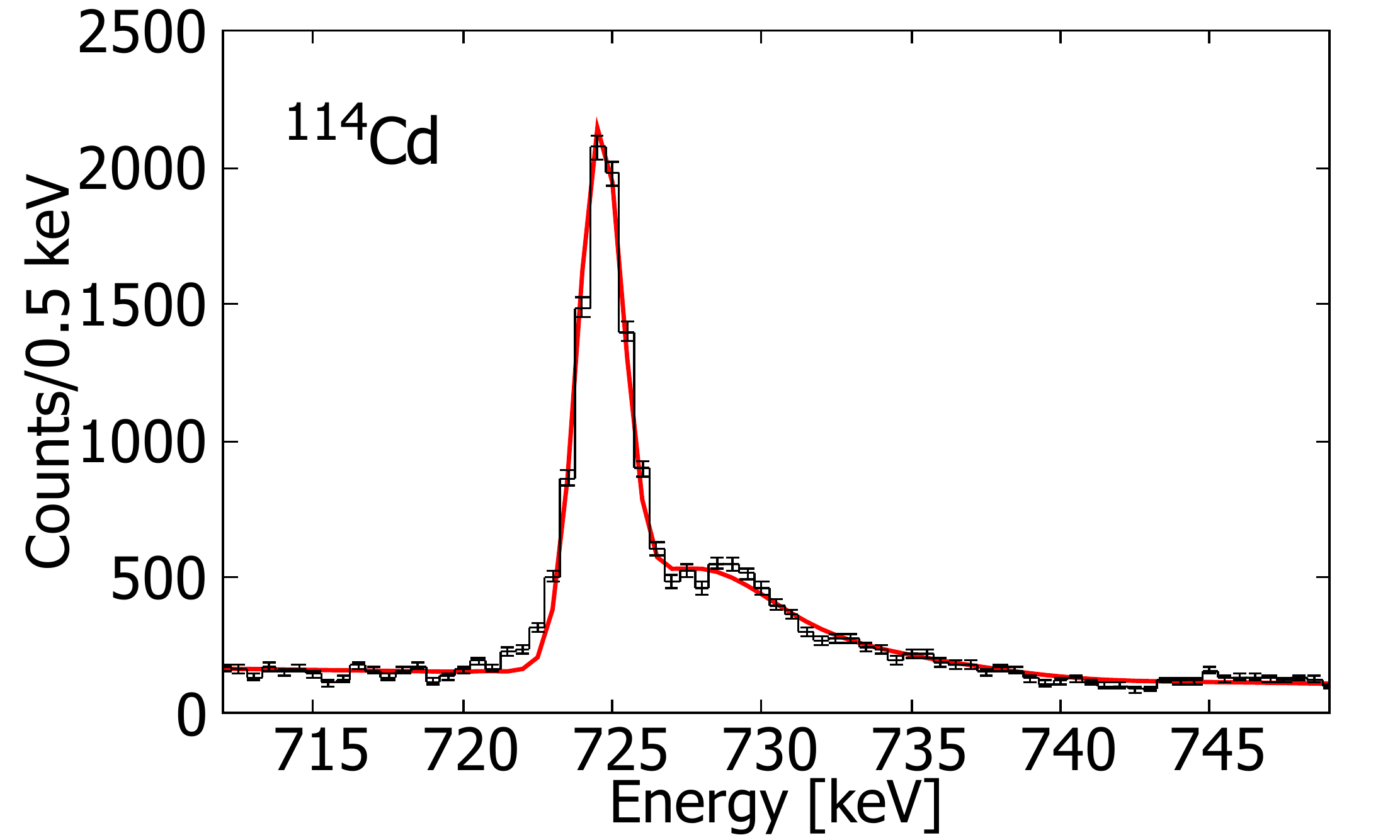}
}
\caption{Doppler-broadened line shape fit to the \mbox{725-keV} transition in $^{114}$Cd. The best-fit line is shown and corresponds to $\tau$=2.26~ps. The previously measured value is $\tau$=2.00(12)~ps~\cite{Blachot2012}. Data from measurements reported in Ref.~\cite{Chamoli2011}.}
\label{DopplerFit725}
\end{figure}

The accuracy of a lifetime determined by the DBLS method is limited by uncertainties in the stopping powers which are typically assumed to be ${\sim}10\%$, but can reach ${\sim}20-30\%$. The complex multilayer targets employed for transient-field $g$-factor measurements add uncertainty by introducing a series of stopping materials each with some uncertainty in its thickness and stopping power. In the present work, Doppler-broadened line shapes were evaluated using the stopping powers of Ziegler~\cite{Ziegler1980} and benchmarked for Cd in iron under the conditions of the experiment by analyzing the transitions depopulating the $4_{1}^{+}$ levels in even Cd isotopes with known lifetime as observed in the data of Chamoli \textit{et al.}~\cite{Chamoli2011}. An example fit to the $4_{1}^{+}\rightarrow2_{1}^{+}$ transition in $^{114}$Cd is shown in Fig.~\ref{DopplerFit725}. The lifetimes extracted from the analysis of the even isotopes differed by up to 30\% from the literature values. We therefore assign a 30\% systematic uncertainty to the extracted lifetimes to cover uncertainties in the stopping power and the complexities of the multilayer target. The statistical uncertainty in the line shape fit was estimated through a chi-squared analysis yielding a $\sim20\%$ uncertainty.

\begin{figure}[t!]
\centerline{
\includegraphics[width=\columnwidth]{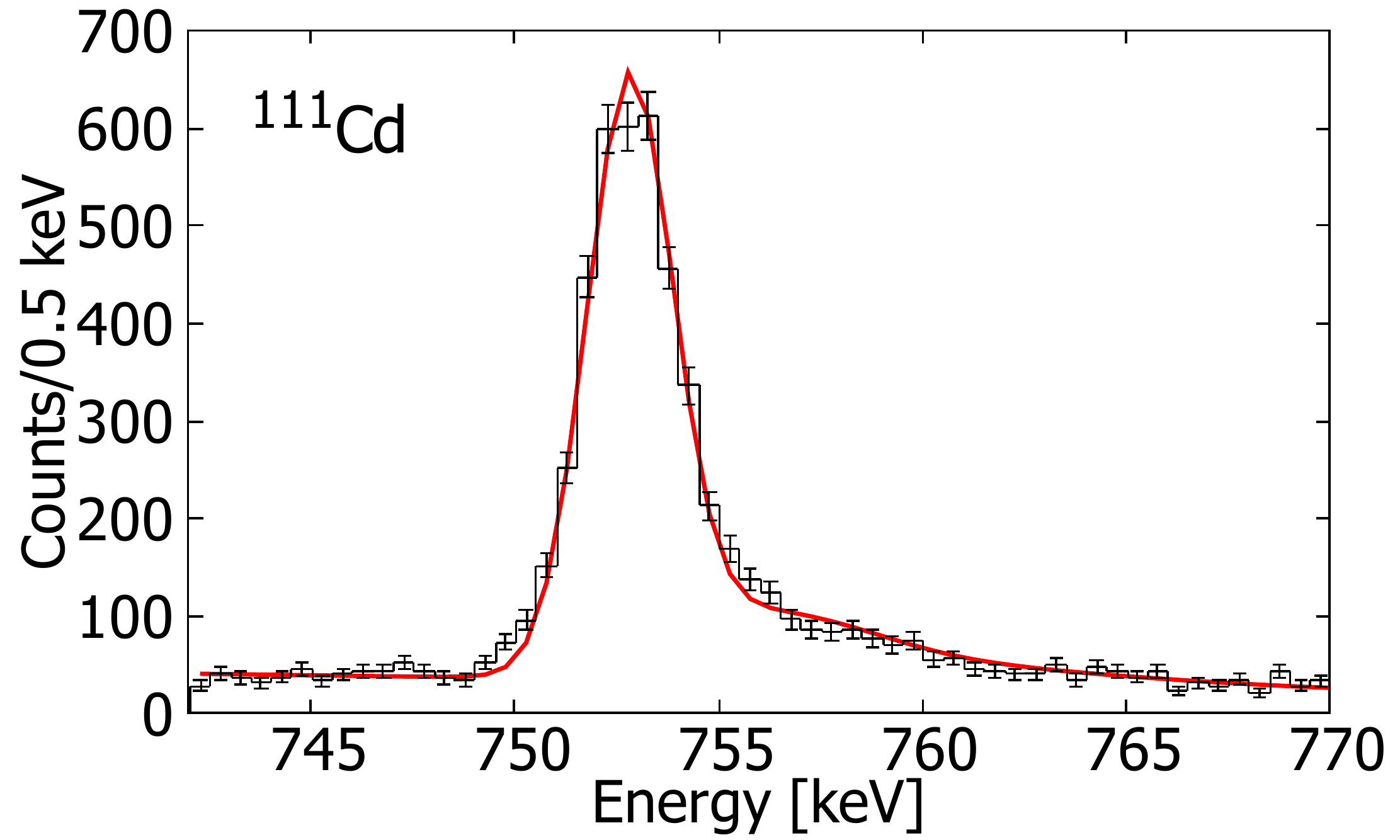}
}
\caption{Doppler-broadened line shape fit to the \mbox{753-keV} transition in $^{111}$Cd. The line shown corresponds to $\tau$=4.0~ps.}
\label{DopplerFit}
\end{figure}

\begin{figure}[t]
\centerline{
\includegraphics[width=\columnwidth]{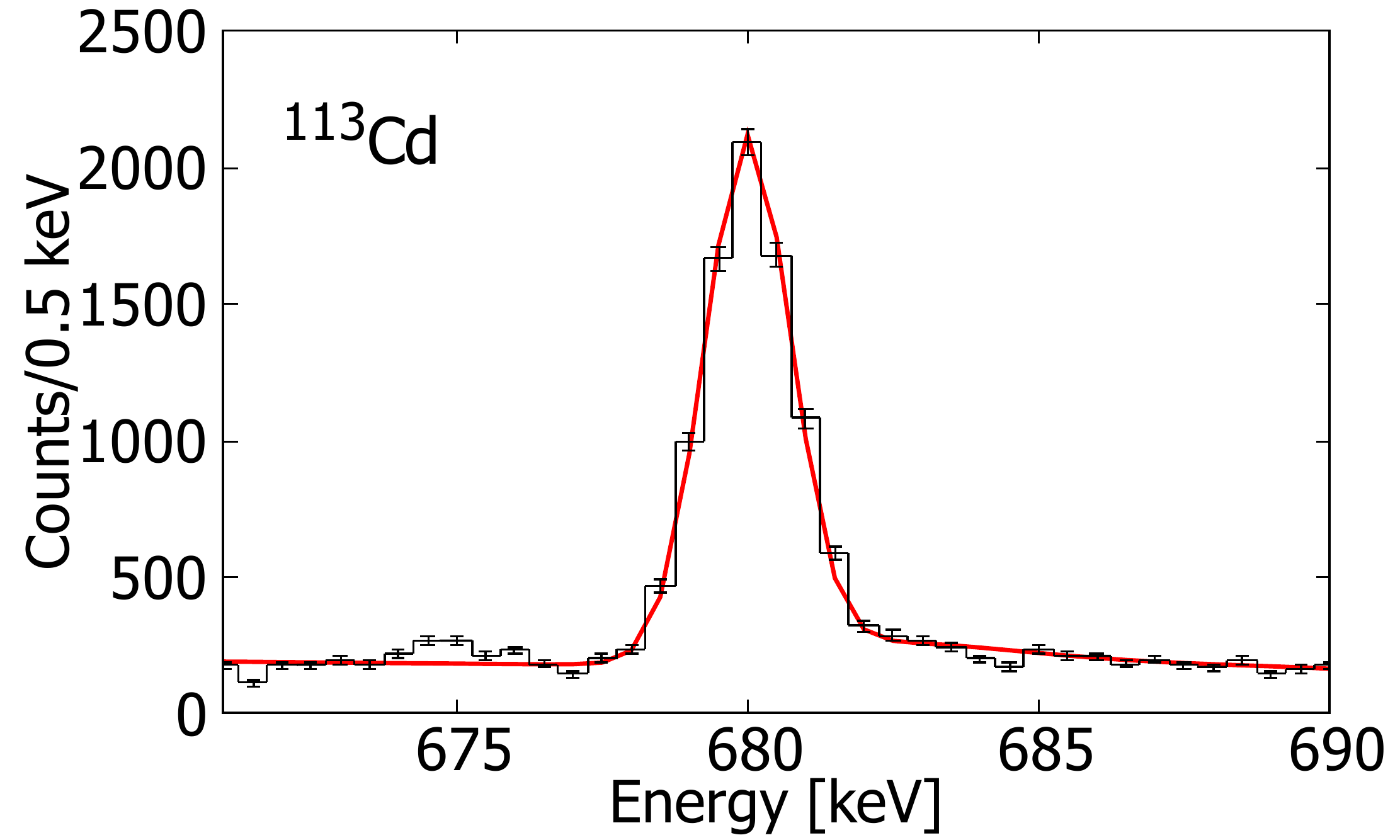}
}
\caption{Doppler-broadened line shape fit to the \mbox{681-keV} transition in $^{113}$Cd. The line shown corresponds to $\tau$=9~ps. Data from measurements reported in Ref.~\cite{Chamoli2011}.}
\label{DopplerFit681}
\end{figure}

The results of the DBLS analysis for the \mbox{753-keV} transition to the ground state in $^{111}$Cd and the \mbox{681-keV} transition to the ground state in $^{113}$Cd are shown in Fig.~\ref{DopplerFit} and Fig.~\ref{DopplerFit681}, respectively. The fitted line shape for the \mbox{753-keV} transition corresponds to a lifetime of 4.0($\pm$1.0~statistical~$\pm$1.2~systematic)~ps and a $B(E2)$ of 11.2($^{+3.7}_{-2.2}$~statistical~$^{+5.8}_{-2.6}$~systematic)~W.u. using the branching ratios obtained in the present work.

As an aside, we note that there is a discrepancy in Nuclear Data Sheets~\cite{Blachot2010} between the reported half-life of the \mbox{681-keV} state of $^{113}$Cd and the $B(E2)$ transition strength measured between this state and the ground state. The reported meanlife is 17(4)~fs while the reported $B(E2)$ strength in the original reference is \mbox{$B(E2\uparrow)=0.070(15)$~$e^{2}b^{2}$}~\cite{Andreev1972}, which corresponds to a meanlife of 8(2)~ps. The lifetime of the \mbox{681-keV} state is evidently too long to be accurately determined by the DBLS method, so only a lower limit could be estimated from the present data. The line shape displayed for the \mbox{681-keV} in Fig.~\ref{DopplerFit681} state corresponds to a lifetime of 9~ps in agreement with the previous Coulomb excitation measurement. Any lifetime greater than this value can be shown to fit the data well and so based on the DBLS fits a limit on the lifetime of $\tau\gtrsim 7$~ps is proposed, consistent with the $B(E2)$ measurements~\cite{McGowan1958,Andreev1972}.

\begin{table*}
  \caption{Experimental \textit{g}~factors in $^{111}$Cd including results of the present transient-field measurements. The measured quantities $\epsilon$, $S$, and $\Delta\theta$ are defined in the text. Adopted values for the first two entries are from the literature. For the remaining entries, measured by the transient-field method, adopted values are weighted averages of present and previous measurements. }
  \centering
  \def\arraystretch{1.5}
  \begin{ruledtabular}
  \begin{tabular}{  c  c  c  c  c  c  c  c  c  }
   $E$ (keV) & $I_{i}^{\pi}$     & $\epsilon\times10^{3}$ & $S$ (rad$^{-1}$) & $\Delta\theta$ (mrad)  &                \multicolumn{4}{c}{$g$~factor}                                  \\ \cline{6-9}
             &                   &                        &                  &                        & Present      & Ref.~\cite{Benczer1989} ~\footnotemark[1] & Ref.~\cite{Stuchbery2016} & Adopted   \\ \hline
       0     & $\frac{1}{2}^{+}$ &   N/A                  &      N/A         &   N/A                  &              &                         &                           & $-$1.189772(2)~\footnotemark[2]   \\
   245.4     & $\frac{5}{2}^{+}$ &   N/A                  &      N/A         &   N/A                  &              &                         &                           & $-$0.306(1)~\footnotemark[2] \\
   342.1     & $\frac{3}{2}^{+}$ & +8(2)                  & -0.18(4)         & -45(13)                & +1.2(4)~\footnotemark[3]      & +0.03(110)              & +0.6(4)                   & $+$0.9(3)    \\
   620.2     & $\frac{5}{2}^{+}$ & +17(4)                 & -1.43(7)         & -12(3)                 & +0.27(7)     & +0.15(6)                & +0.25(7)                  & $+$0.22(4)   \\
   752.8     & $\frac{5}{2}^{+}$ & +31(11)                & -1.43(10)        & -22(8)                 & +0.5(2)      &                         &                           & $+$0.5(2)    \\
  \end{tabular}
  \end{ruledtabular}
  \label{gfactable}
  \footnotetext[1]{Results are from Table II of Ref.~\cite{Benczer1989} which also reports $g(2^+_1 ; ^{110}{\rm Cd}) = +0.382(17)$, consistent with $g(2^+_1 ; ^{110}{\rm Cd}) = +0.407(29)$ obtained subsequently in Ref.~\cite{Chamoli2011}. These results are therefore consistent with the transient-field strength calibration adopted here. The final values reported in Ref.~\cite{Benczer1989} renormalize the $g$~factors to a smaller reference value of $g(2^+_1 ; ^{110}{\rm Cd}) = +0.273(35)$, which is not compatible with the subsequent work.}
  \footnotetext[2]{From Ref.~\cite{Blachot2009}. See also Ref.~\cite{Raghavan1989}. }
  \footnotetext[3]{The observed precession effect for this state includes a small contribution due to feeding from the higher excited states, which is taken into account \cite{Stuchbery2007} in the evaluation of the $g$~factor.}
\end{table*}

\begin{ruledtabular}
\begin{table}[t!]
  \caption{Measured lifetimes from DBLS analysis.}
 \centering
  \def\arraystretch{1.5}
  \begin{tabular}{  c  c  c  c  }
  Nuclide    &  $E_{i}$   & $I_{i}$            &    $\tau$  \\
             &   (keV)    &                    &     (ps)   \\ \hline
  $^{111}$Cd &     753    & $\frac{5}{2}^{+}$  &    4.0(16)  \\
  $^{113}$Cd &     681    & $\frac{3}{2}^{+}$  &    $\gtrsim 7$  \\
  \end{tabular}
  \label{MeasuredLifetimes}
\end{table}
\end{ruledtabular}

\section{Calculations and discussion} \label{DiscussionSection}

In the previous Coulomb excitation study of $^{111,113}$Cd by Stuchbery~\textit{et al.}~\cite{Stuchbery2016} a strongly excited $3/2^{+}$ state was observed at 681~keV in $^{113}$Cd. It was therefore suggested that one of the two states appearing in Nuclear Data Sheets~\cite{Blachot2009} around 755~keV was a state equivalent to this observed \mbox{681-keV} $3/2^{+}$ state in $^{113}$Cd. However, the present experiments have shown that the \mbox{755-keV} state does not exist, and the observation of the $E2$ angular correlation for the ground-state transition from the \mbox{753-keV} state confirms the previously assigned $5/2^{+}$ spin and parity; the \mbox{753-keV} state is clearly not the equivalent state to the $3/2^{+}$ state at \mbox{681 keV} in $^{113}$Cd.

In this section the structure of $^{111}$Cd is discussed in terms of excitation energies, $g$~factors, and reduced transition strengths. In Sec.~\ref{PVSection}, experimental results are compared with calculations performed in the particle-vibration model~\cite{Bohr1953,Choudhury1954} with successes and limitations discussed. Following the previous success~\cite{Stuchbery2016,Wang1988} in describing aspects of the low-lying structure of $^{111,113}$Cd in the particle-rotor model, a description of the levels in $^{111}$Cd excited in the present work is given in terms of Nilsson-scheme based rotational bands in Sec.~\ref{PRSection}. In Sec.~\ref{SMSection} the large-scale shell-model calculations are described and comparisons made with experiment.

\subsection{Particle\hspace{0.05 em}-\negthinspace{}Vibration model} \label{PVSection}

\subsubsection{Calculations}

The particle-vibration model allows calculation of the energy and electromagnetic properties of states and transitions in odd-$A$ vibrational nuclei~\cite{Bohr1953}. The basis states for these calculations are taken to be coupled eigenstates of the collective and quasiparticle Hamiltonians. Interactions between the quasiparticle and collective states are accounted for by an interaction term in the Hamiltonian,
\begin{equation}
\hat{H}_{\rm{int}}=-(\frac{1}{5}\pi)^{\frac{1}{2}}\xi\hbar\omega\sum_{i, \mu}(b^{\mu}+(-1)^{\mu}b_{\mu}^{\dagger})Y_{2\mu}(\hat{r}_{i}),
\end{equation}
where $\mu$ is the $z$-component of the vibrational angular-momentum, $\xi$ is a dimensionless variable controlling the strength of the coupling between the particle and the core, $\hbar\omega$ is the core excitation energy, $b^{\mu}$ is the annihilation operator and $b^{\dagger}_{\mu}$ is the creation operator for the core quadrupole vibrations, and $Y_{2\mu}$ is a spherical harmonic. Up to two core phonons were included in the calculations.

Following usual procedures to define model parameters, the energy spectrum below $\sim1$~MeV excitation in the nucleus of interest was reproduced first and then the electromagnetic observables were examined. The energy spectrum is determined by single-particle energies, $\hbar\omega$, and the strength of the coupling parameter $\xi$. For $^{111}$Cd the core excitation energy was initially set to 610~keV, midway between E$_{2^{+}}$ in $^{110}$Cd and $^{112}$Cd, before varying parameters. The single-particle energies were initially set from the excitation energies in $^{111}$Cd and then allowed to vary, along with $\xi$ and $\hbar\omega$, to yield a ``best" fit to experimental levels. The $E2$ transition strengths stemming from the core vibration scale as $\eta^{2}$, where $\eta=\frac{3}{4\pi}Ze R^{2}_{0}(\hbar\omega/2C_{\rm{Stiff}})^{1/2}$ and $C_{\rm{Stiff}}$ is the surface stiffness. This parameter can be set from the $B(E2)$ of the core using $B(E2;2\rightarrow0)=\eta^{2}$ with $\hbar\omega=\rm{E}_{2^{+}}$.

\begin{figure}[t]
\centerline{
\includegraphics[width=1.5\columnwidth,angle=-90]{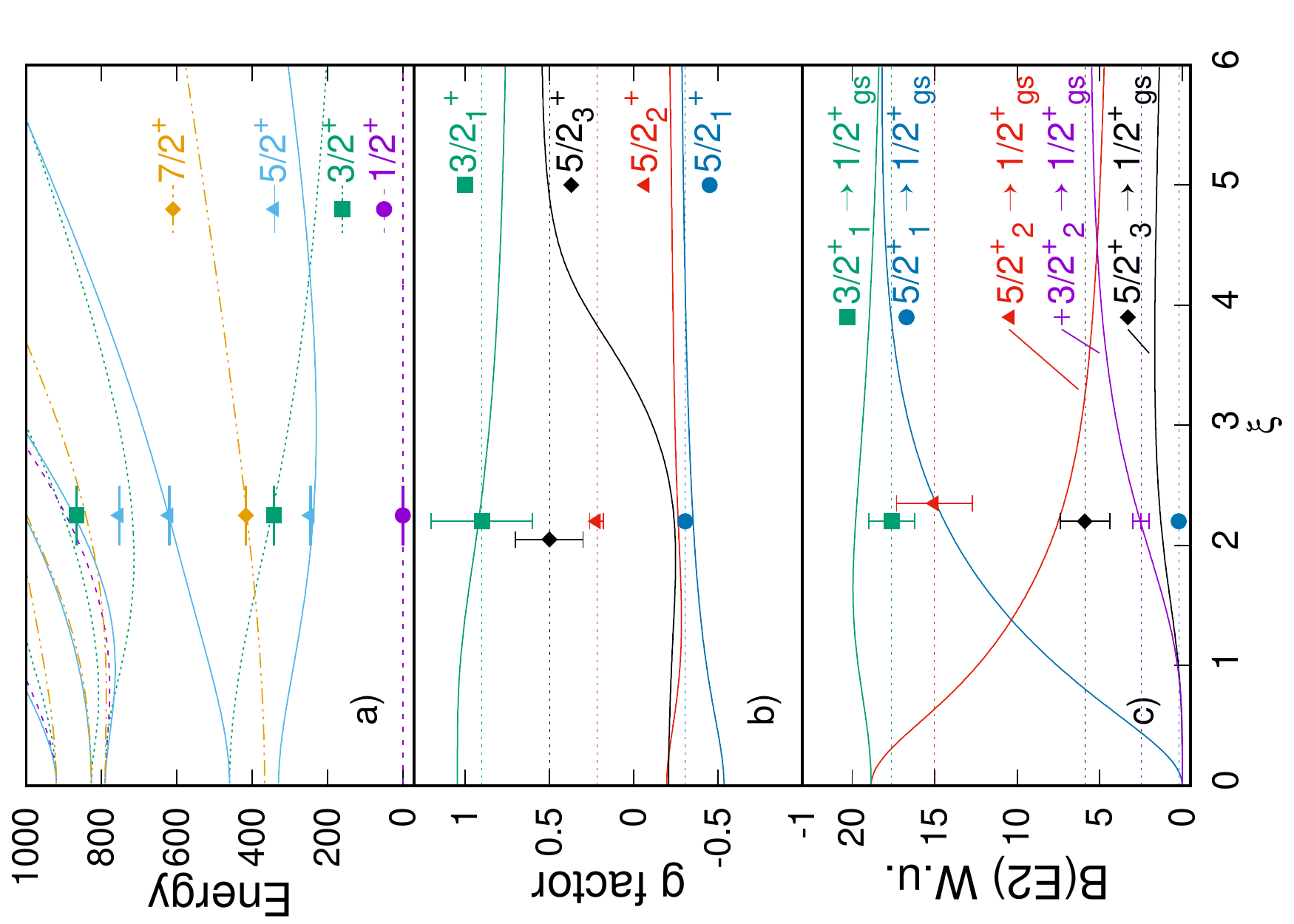}
}
\caption{A comparison between the experimental and theoretical observables as a function of particle-vibration coupling parameter. Experimental data points are placed at the fitted particle-vibration coupling parameter value of $\xi=2.2$ (with some small displacements for clarity of presentation). a) Level energies. Horizontal bars show experimental level energies; lines show the calculations. b) Adopted $g$~factors. Dotted lines show experimental $g$~factors. Solid lines are calculated values. c) Reduced transition strengths for the transitions to the ground state in Weisskopf units. Solid lines are calculated values. Data from Ref.~\cite{Blachot2009}.}
\label{PVlevels}
\end{figure}

The energy spectrum was reproduced well with \mbox{$\xi=2.2$} as seen in Fig.~\ref{PVlevels}~(a). This coupling parameter is similar to that found in previous studies in odd-$A$ Mo and $^{125}$Sb~\cite{Choudhury1969,Heyde1967}, where $\xi\sim1.5-3$ and 2.25, respectively, were used and so the present value is reasonable in this region. The best-fit model parameters are summarized in Table~\ref{PVParameters}. Figure~\ref{PVlevels} shows the dependence of the energy levels, $g$~factors, and $B(E2)$ values as a function of the particle-vibration coupling strength $\xi$, with other parameters held fixed at the values given in Table~\ref{PVParameters}. Experimental values are plotted at the adopted value of $\xi=2.2$.

\begin{ruledtabular}
\begin{table}
  \caption{Particle-vibration model parameters. }
 \centering
  \def\arraystretch{1.5}
  \begin{tabular}{  c  c  c  c  c  c  c  c  }
    $\xi$   &   $\hbar\omega$    & $E_{s_{1/2}}$  & $E_{d_{5/2}}$  & $E_{f_{7/2}}$  &  $E_{d_{3/2}}$   & $g_{R}$  & $C_{\rm{Stiff}}$ \\
            &       (keV)        &   (keV)        &   (keV)        &    (keV)       &      (keV)       &          &   (MeV)          \\ \hline
    2.2    &        440         &      0         &    330         &     367        &      1100        &   0.432  &   55.9           \\
  \end{tabular}
  \label{PVParameters}
\end{table}
\end{ruledtabular}

\renewcommand{\arraystretch}{1.5}

\begin{ruledtabular}
\begin{table}
\caption{Dominant wavefunction components in particle-vibration calculations at $\xi$=2.2. Subscripts in the configuration denote the phonon number of the core state.}
\begin{tabular}{  c  c  c  c  c  c  c  }
  $I_{n}^{\pi}$ & $E_{\rm{exp}}$ & $E_{\rm{theory}}$ & \multicolumn{4}{c}{Configuration} \\
                &(keV)&(keV)&  \multicolumn{4}{c}{Amplitude}        \\
  \hline

\renewcommand{\arraystretch}{1.05}

  $\frac{1}{2}_{gs}^{+}$ & 0 & 0 & \begin{tabular}{c} $s_{\frac{1}{2}}$\\ 0.880 \end{tabular} & \begin{tabular}{c} $d_{\frac{5}{2}}\otimes2^{+}_{1}$\\ 0.395 \end{tabular} & \begin{tabular}{c} $d_{\frac{3}{2}}\otimes2^{+}_{1}$\\ 0.199 \end{tabular} & \begin{tabular}{c} $s_{\frac{1}{2}}\otimes0_{2}^{+}$\\ 0.118 \end{tabular} \\

  $\frac{3}{2}_{1}^{+}$ & 342 & 352 & \begin{tabular}{c} $s_{\frac{1}{2}}\otimes2^{+}_{1}$\\ 0.793 \end{tabular} & \begin{tabular}{c} $d_{\frac{3}{2}}$\\ -0.324 \end{tabular} & \begin{tabular}{c} $d_{\frac{5}{2}}\otimes4^{+}_{2}$\\ 0.320 \end{tabular} & \begin{tabular}{c} $g_{\frac{7}{2}}\otimes4^{+}_{2}$\\ -0.270 \end{tabular} \\

  $\frac{3}{2}_{2}^{+}$ & 867 & 717 & \begin{tabular}{c} $g_{\frac{7}{2}}\otimes2^{+}_{1}$\\ 0.605 \end{tabular} & \begin{tabular}{c} $s_{\frac{1}{2}}\otimes2_{2}^{+}$\\ 0.424 \end{tabular} & \begin{tabular}{c} $d_{\frac{3}{2}}$\\ 0.346 \end{tabular} & \begin{tabular}{c} $d_{\frac{5}{2}}\otimes2^{+}_{1}$\\ 0.333 \end{tabular} \\

  $\frac{5}{2}_{1}^{+}$ & 245 & 239 & \begin{tabular}{c} $d_{\frac{5}{2}}$\\ -0.670 \end{tabular} & \begin{tabular}{c} $s_{\frac{1}{2}}\otimes2^{+}_{1}$\\ -0.609 \end{tabular} & \begin{tabular}{c} $d_{\frac{5}{2}}\otimes2^{+}_{1}$\\ 0.294 \end{tabular} & \begin{tabular}{c} $d_{\frac{5}{2}}\otimes4^{+}_2$\\ -0.149 \end{tabular} \\

  $\frac{5}{2}_{2}^{+}$ & 620 & 618 & \begin{tabular}{c} $s_{\frac{1}{2}}\otimes2^{+}_{1}$\\ 0.652 \end{tabular} & \begin{tabular}{c} $d_{\frac{5}{2}}\otimes2^{+}_{1}$\\ 0.440 \end{tabular} & \begin{tabular}{c} $d_{\frac{5}{2}}$\\ -0.420 \end{tabular} & \begin{tabular}{c} $d_{\frac{5}{2}}\otimes2_{2}^{+}$\\ 0.274 \end{tabular} \\

  $\frac{5}{2}_{3}^{+}$ & 753 & 865 & \begin{tabular}{c} $d_{\frac{5}{2}}\otimes2^{+}_{1}$\\ 0.559 \end{tabular} & \begin{tabular}{c} $s_{\frac{1}{2}}\otimes2_{2}^{+}$\\ 0.523 \end{tabular} & \begin{tabular}{c} $d_{\frac{5}{2}}$\\ 0.518 \end{tabular} & \begin{tabular}{c} $d_{\frac{5}{2}}\otimes4^{+}_{2}$\\ -0.220 \end{tabular} \\

\end{tabular}
\label{WavefunctionTable}
\end{table}
\end{ruledtabular}

\renewcommand{\arraystretch}{1}

\subsubsection{Results and discussion}

The dominant wavefunction components of the low-lying excited states are given in Table~\ref{WavefunctionTable} for the calculation at $\xi=2.2$. Overall, it is possible to obtain a close energy fit for low excitation states; however, at higher energies, where the predicted density of states increases, the calculated energy levels show little resemblance to experiment.

The observed $3/2_{1}^{+}$ state is calculated to be largely composed of the core-phonon excitation coupled to the $s_{1/2}$ neutron. The next expected $3/2^{+}$ state is the $d_{5/2}\otimes2^{+}$ configuration in the weak-coupling limit, $\xi=0$. As coupling strength increases, however, the second $3/2^{+}$ state undergoes strong configuration mixing, gaining stronger $g_{7/2}\otimes2^{+}$ and $s_{1/2}\otimes2^{+}$ components. The resulting $3/2^{+}$ state appears in the calculations around $\sim720$~keV, however, the second excited $3/2^{+}$ state observed experimentally in the current work occurs higher in energy (867~keV). The non-observation of a second low-lying collective $3/2^{+}$~state is a challenge to the particle-vibration model.

The $g$~factors for the ground state, and $5/2_{2}^{+}$ excited state have previously been shown to be inconsistent with the particle-vibration model~\cite{Stuchbery2016}. Continuing the same trend as for the $5/2_{2}^{+}$ state, the measured $g$~factor for the $5/2_{3}^{+}$ state measured here for the first time is positive while the predicted $g$~factor is negative. It is possible to force a positive $g$~factor for the \mbox{753-keV} $5/2^{+}_{3}$ state in the model by increasing mixing from either the $d_{3/2}$ single-particle orbit, or from the $g_{7/2}$ single-particle orbit. Each of these components of the $5/2_{3}^{+}$ state increases as the particle-core coupling increases, with the $g_{7/2}$ component becoming dominant. However, the reasonably good description of the energy levels is lost.

A comparison of the experimental and calculated transition strengths is shown in panel c) of Fig.~\ref{PVlevels}. These calculations were performed with a stiffness parameter based on the transition strength of the even core. The relative magnitudes of $E2$ reduced transition strengths are reasonably well reproduced with the exception of the decay of the first-excited $5/2^{+}_{1}$ state, which experimentally is isomeric, and is most naturally ascribed to a dominant $d_{5/2}\otimes0^{+}$ configuration. This interpretation is supported by the measured $g$~factor, which has the sign and about half the magnitude of the Schmidt value. Increasing the particle-coupling parameter increases the proportion of $s_{1/2}\otimes2^{+}$ in the wavefunction of the $5/2^{+}_{1}$ state. The consequent change in collectivity causes a drastic increase in the strength of the transition to the ground state, which is inconsistent with the isomeric character of the state. The natural explanation of the $g$~factor and the lifetime of the $5/2_{1}$ state remains that it is predominantly a $\nu d_{5/2}$ single-particle excitation.

The Cd isotopes have recently been studied utilising an effective field theory model~\cite{Coello2016}. The $g$-factor predictions made in that paper are consistent with experiment for the ground state and first-excited $3/2_{1}^{+}$ state, however still incorrectly predict the sign and magnitude of the $g$~factor for the second-excited $5/2_{2}^{+}$ state. No prediction was made for the newly-measured $5/2^{+}_{3}$ state. As noted by the authors, there is no simple relationship between their model parameters and the familiar Schmidt single-particle moments. Thus, the overall fair agreement of the model $g$~factors with experiment benefits from the use of the measured $g$~factors to set the relevant low-energy constants (i.e.~model parameters).

\subsection{Particle-Rotor model} \label{PRSection}

As the particle-vibration model does not readily explain the observed nuclear structure of $^{111}$Cd, attention was given to the particle-rotor model. Previous studies~\cite{Wang1988,Zeghib2007} of odd-mass Cd isotopes have found that $^{111,113}$Cd are well described by a particle-rotor description with weak deformation and a variable moment of inertia.

\begin{figure}[ht]
\includegraphics[width=0.9\columnwidth]{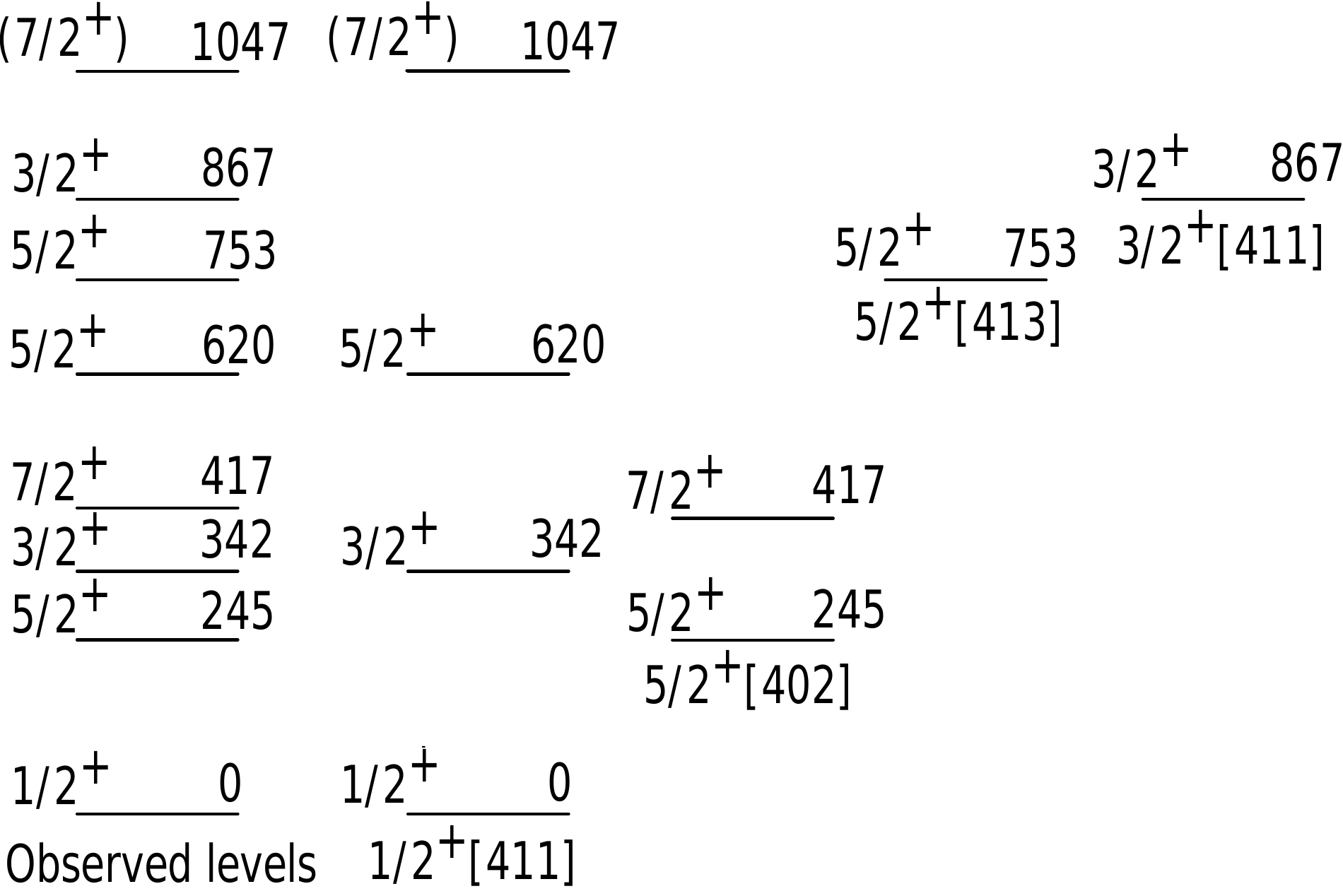}
\caption{Arrangement of the observed low-excitation energy, positive-parity levels in $^{111}$Cd into a particle-rotor level scheme.}
\label{PRDecomposition}
\end{figure}

A rigorous fitting to the observed levels was not performed in the present work as the particle-rotor model in this weakly deformed region is quite parameter dependent. Instead, a heuristic interpretation of the observed states is given. The classification of states discussed in this section is shown in Fig.~\ref{PRDecomposition} where the states are labeled using the terminology of the strong-coupling limit of the particle-rotor model (cf. Ref.~\cite{Wang1988}). This nomenclature should not be taken to imply that the strong coupling limit is applicable at small deformations. The perspective here is that a particle-rotor description may give useful insights due to Lawson's observation that Nilsson wavefunctions at small deformation provide a good approximation to the shell model~\cite{Lawson1980}.

\subsubsection{$g$~factors}

\begin{figure}[t!]
\includegraphics[width=\columnwidth]{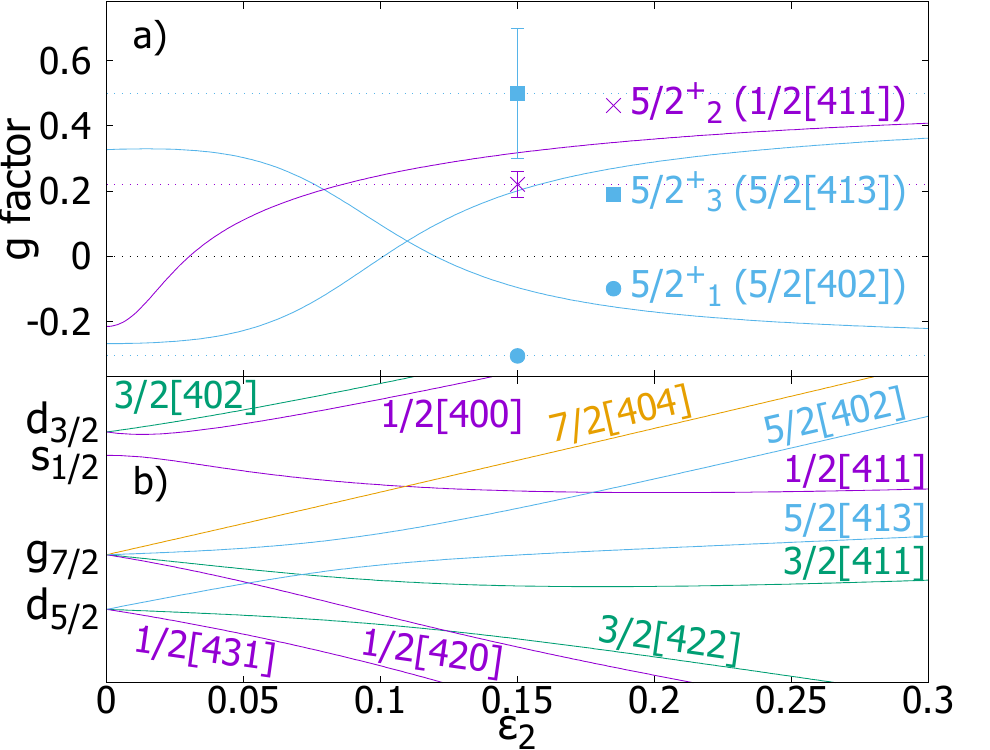}
\caption{a) Nilsson model calculations of \textit{g}~factors as a function of quadrupole deformation $\epsilon_{2}$. The theoretical $g$~factors of each state show a marked change with deformation moving towards the experimental values as deformation increases. b) Positive-parity Nilsson scheme calculated for $^{111}$Cd.}
\label{NilssonStructuregfac}
\end{figure}

It has previously been shown that the $g$~factors of the ground state and low-lying excited states in $^{111}$Cd can be interpreted through mixing induced by a small quadrupole deformation~\cite{Stuchbery2016}. In particular, the avoided crossing between the $1/2^{+}[400]$ and $1/2^{+}[411]$ Nilsson orbits shown at small deformations ($\epsilon_{2}<0.05$) in Fig.~\ref{NilssonStructuregfac} is associated with a change in the composition of the $1/2^{+}[411]$ orbit from primarily $s_{1/2}$ parentage to primarily $d_{3/2}$ parentage, which strongly affects the $g$~factor. Similarly, the measured $g$~factors for the intrinsic $5/2^{+}$ band-head states may be explained by an exchange of odd-particle character at avoided crossings as deformation increases, as indicated in Fig.~\ref{NilssonStructuregfac}. More specifically, as the energies of two Nilsson orbits of the same spin and parity become close, mixing increases and the states are seen to move apart or avoid each other. As the deformation increases in the $^{111}$Cd calculations, the $5/2^{+}[402]$ Nilsson orbit with $g_{7/2}$ parentage undergoes an avoided crossing with the $5/2^{+}[413]$ single-particle orbit of $d_{5/2}$ parentage allowing significant sharing of the $d_{5/2}$ character.

In this case, the change in nature of the $5/2^{+}[402]$ orbit near $\epsilon_{2}=0.1$ causes the $g$~factor to change sign. The $5/2_{1}^{+}$ state in $^{111}$Cd has a previously measured $g$~factor of $-0.306(1)$~\cite{Blachot2009}; this sign is consistent with the Schmidt value for a $d_{5/2}$ state. The newly measured $g$~factor for the $5/2_{3}^{+}$ is $+0.5(2)$ suggesting it has a large $g_{7/2}$ or $d_{3/2}$ component. Although agreement is not perfect, the particle-rotor model $g$~factors of each of the excited $5/2^{+}$ states moves toward the experimental values as deformation increases.

The $M1$ transition strengths predicted by the particle-rotor model are proportional to $(g_{K}-g_{R})^{2}$, where $g_{R}$ is the rotational $g$~factor and $g_{K}$ is the intrinsic single-particle $g$~factor. As deformation increases the intrinsic $g$~factor for the $1/2^{+}[411]$ ground state increases and is approximately equal to the rotational $g$~factor at $\epsilon_{2}\sim0.1$, consistent with the deformation implied by measured quadrupole moments. This cancellation can produce small $M1$ transition strengths as seen in experiment. However, the Nilsson model $g_{K}$ values are highly deformation dependent and therefore $g_{K}-g_{R}\approx0$ is not assured.

\subsubsection{Ground-state band}

In the particle-rotor description, the ground-state band is built on the $1/2^{+}[411]$ Nilsson orbit. At small deformation the strong mixing between the $s_{1/2}$ and $d_{3/2}$ orbits acts both to lower the energy of the $1/2^{+}[411]$ Nilsson orbit of $s_{1/2}$ parentage and increase the $g$~factor of the ground state toward the observed value~\cite{Stuchbery2016}. The band built on this state can be identified through observation of $E2$ transition strengths. The transitions to the ground state from the $3/2_{1}^{+}$ and $5/2_{2}^{+}$ states are enhanced, confirming a collective structure; however, as noted previously~\cite{Stuchbery2016}, the relative strength of these transitions is the same in the particle-rotor and weak-coupling descriptions. The observation of the higher-energy (tentatively) $7/2_{2}^{+}$ state at 1047~keV in our work is suggestive of a large $B(E2)$ value. The apparent collectivity of this state, and the fact that it is about at the expected excitation energy, suggest that it could be part of the ground-state band. However, further spectroscopic measurements are required to make a firm assignment.

\subsubsection{The \mbox{245-keV} $\frac{5}{2}^{+}$ and \mbox{417-keV} $\frac{7}{2}^{+}$ states}

The $B(E2)$ value of 0.22 W.u. for the transition from the \mbox{245-keV} state to the ground state~\cite{Blachot2009} is suggestive that this is a single-particle state. The negative $g$~factor then identifies the state as having a large $d_{5/2}$ component, consistent with the observation of an avoided crossing between the $5/2^{+}[402]$ and $5/2^{+}[413]$ Nilsson orbits. The reported $B(E2)=23(6)$~W.u.~\cite{Blachot2009} from the $7/2_{1}^{+}$ state to the \mbox{245-keV} $5/2_{1}^{+}$ state suggests that the $7/2_{1}^{+}$ state has significant collective character and so may be a part of a band built on the $5/2_{1}^{+}$ level, which would be the head of the $5/2^{+}[402]$ band. As seen in Fig.~\ref{NilssonStructuregfac}, a $7/2^{+}$ single-particle state corresponding to the $7/2^{+}[404]$ Nilsson orbit is also expected at around this energy but a collective $E2$ transition to the $5/2_{1}^{+}$ \mbox{245-keV} state is then not expected.

\subsubsection{The \mbox{753-keV} $\frac{5}{2}^{+}$ state}

With the confirmation that only one strongly Coulomb excited state appears around $\sim$750~keV excitation energy and that the spin-parity is $5/2^{+}$, the structure of this state must be identified. Naively, one would expect a single-particle state of this spin and parity to exist corresponding to the $5/2^{+}[413]$ Nilsson orbit. The two previously reported $B(E2){\downarrow}$ strengths for the transition from the $5/2_{3}^{+}$ state to the ground state correspond to disparate mean lifetimes of $\sim$14-17~ps~\cite{Singh1985}, or $9(2)$~ps~\cite{McDonald1968}. The line shape for this transition has a prominent Doppler tail, however the lifetime for the state is at the limit of the DBLS technique and therefore cannot be determined reliably in the present work. The lifetime obtained here (4.0$\pm1.6$~ps) favors the $B(E2)$ from Ref.~\cite{McDonald1968} and suggests a collective transition. There are several low-lying $5/2^{+}$ states in this nucleus that can mix. The observation of strong branches from the $5/2_{3}^{+}$ state to the $1/2_{\rm{g.s.}}^{+}$, $5/2_{1}^{+}$, and $7/2_{1}^{+}$ states suggests significant configuration mixing. More detailed spectroscopy is required to fully understand this state.

\subsubsection{The \mbox{867-keV} $\frac{3}{2}^{+}$ state}

The previous measurement of the \mbox{$B(E2;3/2_{2}^{+}\rightarrow1/2_{\rm{g.s.}}^{+})$} value for the \mbox{867-keV} transition~\cite{Blachot2009} is 2.5(5)~W.u. and is consistent with a single-particle state and so it has been identified with the $3/2^{+}[411]$ Nilsson orbit.

\subsubsection{Limitations of the particle-rotor model}

More detailed calculations away from the rigid-rotor model begin to show flaws in the particle-rotor description. In particular, it is difficult to reproduce the energy spectrum of $^{111}$Cd accurately. This is possibly due to the nucleus being soft, with deformation increasing with the rotational angular momentum. An attempt was made to include a variable moment of inertia into the particle-rotor calculations. While improved energy fits were possible, a simultaneous description of electromagnetic observables became harder to achieve. For this reason no attempt has been made to give a detailed comparison of excitation energy, $B(E2)$, or $B(M1)$ with particle-rotor calculations.

\subsection{Shell-model calculations} \label{SMSection}

\subsubsection{Model space and interactions} \label{ModelSpace}

Large-scale shell-model calculations were performed using the $M$-scheme code KSHELL~\cite{Shimizu2013}. The SR88MHJM Hamiltonian~\cite{Kavatsyuk2007,Faestermann2013,Park2017,Yordanov2018} employed for these calculations assumes an inert $^{88}$Sr core and includes orbits up to $Z=50$ and $N=82$. Specifically, the model space is $\pi(p_{\frac{1}{2}},g_{\frac{9}{2}})\nu(d_{\frac{5}{2}},g_{\frac{7}{2}},d_{\frac{3}{2}},s_{\frac{1}{2}},h_{\frac{11}{2}})$. The interaction is based on the CD-Bonn potential with renormalization of the $G$-matrix~\cite{Machleidt1996,HjorthJensen1995,HjorthJensen3000}. Two body matrix elements were then adjusted to reproduce experimental energies in the tin isotopes~\cite{Grawe3000}. The name of the interaction, SR88MHJM, therefore refers to the $^{88}$Sr core and indicates that modifications (M) were made to the original Hamiltonian provided by Morten Hjorth-Jensen (MHJ). For the evaluation of electromagnetic transition rates, the effective charges were $e_{\nu} = e$ and $e_{\pi} = 1.7e$ and spin $g$~factors were quenched to 0.7 of the bare value. The $g$~factors were evaluated for both the bare operator and with the effective spin $g$~factors quenched as for the transition rate calculations.

\subsubsection{Results and discussion} \label{SMCalculations}

The excitation energies and $g$~factors are compared with experiment in Table~\ref{ShellModelResults}. The RMS-deviation between the theoretical and experimental energies is 69~keV, an excellent level of agreement between theory and experiment.

Overall, the shell-model $g$~factors are in good agreement with experiment. The signs of the $g$~factors are reproduced in all measured cases and generally the experimental magnitude is within the range spanned by calculations with free and effective $g_{s}$ values. Thus, the shell-model calculations better describe the $g$~factors than the particle-vibration model, and are on a par with the particle-rotor description.

In fact the $g$~factors in all measured cases agree well with the shell model, except for the $5/2^{+}_{3}$~state which is underpredicted by two standard deviations. This difference between theory and experiment may signal deficiencies in the wavefunction. In the work of Yordanov \textit{et al.}~\cite{Yordanov2018} on $^{101-109}$Cd the $g$~factors of the first $5/2^{+}$ states were compared to values calculated with the SR88MHJM Hamiltonian up to $^{111}$Cd taking the same effective $g_{s}$ values as used here. Thus the value obtained for $g(5/2^{+}_{1})$ is the same as the value reported in the 5$^{\rm{th}}$ column of Table~\ref{ShellModelResults}. The $g(5/2^{+}_{1})$ values were found by Yordanov \textit{et al.} to be systematically overpredicted toward the mid-shell (see also $5/2^{+}_{1}$ in Table~\ref{ShellModelResults}), which may be attributed to increasing effects from an incomplete basis space.

\begingroup
\squeezetable
\begin{ruledtabular}
\begin{table}
  \caption{Level energies and $g$-factors in $^{111}$Cd based on shell-model calculations with the SR88MHJM Hamiltonian. }
 \centering
  \def\arraystretch{1.5}
  \begin{tabular}{  c  c  c  c  c  c  }
$E_{\rm{exp}}$~(keV) & $E_{\rm{theory}}$~(keV)&   $I_{n}^{\pi}$               & \multicolumn{3}{c}{$g$~factor}   \\ \cline{4-6}
                     &                        &                               &     free~\footnotemark[1]  &   effective~\footnotemark[2] &      exp~\footnotemark[3]   \\ \hline
            0        &     0                  & $\frac{1}{2}_{1}^{+}$         &    $-$1.295       &      $-$0.858     &     $-$1.189772(2)   \\
          245        &   171                  & $\frac{5}{2}_{1}^{+}$         &    $-$0.370       &      $-$0.244     &     $-$0.306(1) \\
          342        &   216                  & $\frac{3}{2}_{1}^{+}$         &     +0.624        &      +0.473       &     +0.9(3)     \\
          396        &   307                  & $\frac{11}{2}_{1}^{-}$        &    $-$0.293        &      $-$0.200     &     $-$0.20093(7)    \\
          417        &   480                  & $\frac{7}{2}_{1}^{+}$         &     +0.299        &      +0.227       &                 \\
          620        &   632                  & $\frac{5}{2}_{2}^{+}$         &     +0.207        &      +0.162       &     +0.22(4)    \\
          753        &   787                  & $\frac{5}{2}_{3}^{+}$         &     +0.098        &      +0.134       &     +0.5(2)     \\
          867        &   808                  & $\frac{3}{2}_{2}^{+}$         &     +0.626        &      +0.459       &                 \\
          854        &   856                  & $\frac{7}{2}_{2}^{+}$         &     +0.119        &      +0.123       &                 \\
  \end{tabular}
  \label{ShellModelResults}
  \footnotetext[1]{Evaluated with free nucleon $g_{s}$ values.}
  \footnotetext[2]{Evaluated with effective nucleon $g_{s}$ values.}
  \footnotetext[3]{Values from Table~\ref{gfactable} and Nuclear Data Sheets~\cite{Blachot2009}.}
\end{table}
\end{ruledtabular}
\endgroup

Particle-hole (proton intruder) excitations, which are excluded from the basis space, could also impact on the $g$~factor of the $5/2^{+}_{3}$ state, which has an excitation energy approaching that of the intruder configurations in the neighboring even Cd isotopes.

Transition rates are compared with experiment in Tables~\ref{SMBM1Table} and \ref{SMBE2Table}. The experimental data in these Tables have been evaluated using the RULER~v.4.1~\cite{Burrows1984} program. Unless otherwise indicated, data are from Ref.~\cite{Blachot2009}.

\begin{ruledtabular}
\begin{table}
  \caption{Results of the $M1$ transition strength shell-model calculations for $^{111}$Cd with the SR88MHJM Hamiltonian (SM) compared with the particle-vibration (PV) model calculations and experiment. Experimental values are from Ref.~\cite{Blachot2009} evaluated using RULER~\cite{Burrows1984}, unless otherwise stated. }
 \centering
  \def\arraystretch{1.5}
  \begin{tabular}{  c  c  c  c  c  }
  $I_{ni}^{\pi}$              &  $I_{nf}^{\pi}$       &   \multicolumn{3}{c}{$B(M1)$ (W.u.)}    \\ \cline{3-5}
                              &                       & Theory (SM)   &Theory (PV)&  Exp        \\ \hline

$\frac{3}{2}_{1}^{+}$         &$\frac{1}{2}_{1}^{+}$  &     0.009     &  0        &   0.020(3)  \\
$\frac{3}{2}_{1}^{+}$         &$\frac{5}{2}_{1}^{+}$  &     0.043     &  0.533    &  0.0170(23) \\

$\frac{7}{2}_{1}^{+}$         &$\frac{5}{2}_{1}^{+}$  &     0.023     &$<$0.001   &   0.036(9)  \\

$\frac{5}{2}_{2}^{+}$         &$\frac{5}{2}_{1}^{+}$  &     0.016     &  0.013    &  0.0011(4)~\footnotemark[1]\\
$\frac{5}{2}_{2}^{+}$         &$\frac{3}{2}_{1}^{+}$  &     0.005     &  0.090    &             \\
$\frac{5}{2}_{2}^{+}$         &$\frac{7}{2}_{1}^{+}$  &   $<$0.001    &$<$0.001   &             \\

$\frac{5}{2}_{3}^{+}$         &$\frac{5}{2}_{1}^{+}$  &   $<$0.001    &  0.012    &             \\
$\frac{5}{2}_{3}^{+}$         &$\frac{3}{2}_{1}^{+}$  &     0.013     &  0.021    & 0.019(8)~\footnotemark[2]\\
$\frac{5}{2}_{3}^{+}$         &$\frac{7}{2}_{1}^{+}$  &     0.007     &  0.003    &             \\
$\frac{5}{2}_{3}^{+}$         &$\frac{5}{2}_{2}^{+}$  &     0.006     &  0.001    &             \\
$\frac{5}{2}_{3}^{+}$         &$\frac{7}{2}_{2}^{+}$  &     0.047     &  0.089    &             \\
$\frac{5}{2}_{3}^{+}$         &$\frac{3}{2}_{2}^{+}$  &     0.026     &  0.006    &             \\

$\frac{3}{2}_{2}^{+}$         &$\frac{1}{2}_{1}^{+}$  &     0.009     &  0.001    &             \\
$\frac{3}{2}_{2}^{+}$         &$\frac{3}{2}_{1}^{+}$  &     0.015     &  0.039    &0.0014($^{+5}_{-6}$)\\
$\frac{3}{2}_{2}^{+}$         &$\frac{5}{2}_{1}^{+}$  &     0.055     &  0.002    &             \\
$\frac{3}{2}_{2}^{+}$         &$\frac{5}{2}_{2}^{+}$  &     0.021     &  0.348    &             \\

$\frac{7}{2}_{2}^{+}$         &$\frac{5}{2}_{1}^{+}$  &     0.053     &  0.010    &             \\
$\frac{7}{2}_{2}^{+}$         &$\frac{7}{2}_{1}^{+}$  &     0.041     &$<$0.001   &             \\
$\frac{7}{2}_{2}^{+}$         &$\frac{5}{2}_{2}^{+}$  &     0.004     &  0.022    &             \\

  \end{tabular}
  \label{SMBM1Table}
  \footnotetext[1]{Transition strength from Ref.~\cite{McGowan1958} evaluated with RULER~\cite{Burrows1984}.}
  \footnotetext[2]{Transition strength calculated from the measured mean lifetime and branching ratios found in the present work. The Nuclear Data Sheet value is 0.0103(24)~W.u. with the new branching ratios.}
\end{table}
\end{ruledtabular}

Table~\ref{SMBM1Table} shows the theoretical and experimental $M1$ transition rates. As may be expected for a weakly collective nucleus, all of the low-excitation $M1$ transitions are weak. The level of agreement between theory and experiment is therefore satisfactory. The $M1$ transition strengths are largely similar between the shell-model calculations and the particle-vibration model calculations. The notable exception is the $3/2^{+}_{1}\rightarrow5/2^{+}_{1}$ transition which is much stronger in the particle-vibration calculation.

\begin{ruledtabular}
\begin{table}
  \caption{Results of the $E2$ transition strength shell-model (SM) calculations for $^{111}$Cd with the SR88MHJM Hamiltonian compared with the particle-vibration (PV) model calculations and experiment. Experimental values are from Ref.~\cite{Blachot2009}, unless otherwise stated. All experimental data were evaluated with RULER~\cite{Burrows1984}. Strong ground-state transitions are in bold.}
 \centering
  \def\arraystretch{1.5}
  \begin{tabular}{  c  c  c  c  c  }
  $I_{ni}^{\pi}$      &  $I_{nf}^{\pi}$       &   \multicolumn{3}{c}{$B(E2)$ (W.u.)}        \\ \cline{3-5}
                      &                       &  Theory (SM)  &  Theory (PV)  &     Exp     \\ \hline

$\frac{5}{2}_{1}^{+}$ &$\frac{1}{2}_{1}^{+}$  &     0.21      &    14.4       & 0.2374(12)  \\

\boldmath$\frac{3}{2}_{1}^{+}$ &\boldmath$\frac{1}{2}_{1}^{+}$& \textbf{14.1} & \textbf{19.8} &\textbf{18(3)}\\
$\frac{3}{2}_{1}^{+}$ &$\frac{5}{2}_{1}^{+}$  &  $<$0.01      &     1.5       &     21(15)  \\

$\frac{7}{2}_{1}^{+}$ &$\frac{5}{2}_{1}^{+}$  &     18.0      &     1.2       &      21(6)  \\
$\frac{7}{2}_{1}^{+}$ &$\frac{3}{2}_{1}^{+}$  &     1.4       &     4.0       &             \\

\boldmath$\frac{5}{2}_{2}^{+}$ &\boldmath$\frac{1}{2}_{1}^{+}$  & \textbf{7.5}  & \textbf{7.5}  &\textbf{15(2)}~\footnotemark[1]  \\
$\frac{5}{2}_{2}^{+}$ &$\frac{5}{2}_{1}^{+}$  &     10.4      &     7.1       & 48(8)~\footnotemark[1]  \\
$\frac{5}{2}_{2}^{+}$ &$\frac{3}{2}_{1}^{+}$  &  $<$0.01      &     1.2       &             \\
$\frac{5}{2}_{2}^{+}$ &$\frac{7}{2}_{1}^{+}$  &     18.1      &     1.3       &             \\

\boldmath$\frac{5}{2}_{3}^{+}$ &\boldmath$\frac{1}{2}_{1}^{+}$  & \textbf{ 15.0}& \textbf{0.42} &\textbf{11(5)}~\footnotemark[2]  \\
$\frac{5}{2}_{3}^{+}$ &$\frac{5}{2}_{1}^{+}$  &     1.8       &    11.8       &             \\
$\frac{5}{2}_{3}^{+}$ &$\frac{3}{2}_{1}^{+}$  &     6.4       &     1.2       &0.23($^{+30}_{-23}$)~\footnotemark[2]\\
$\frac{5}{2}_{3}^{+}$ &$\frac{7}{2}_{1}^{+}$  &     1.8       &    0.67       &             \\
$\frac{5}{2}_{3}^{+}$ &$\frac{5}{2}_{2}^{+}$  &     10.4      &     1.2       &             \\
$\frac{5}{2}_{3}^{+}$ &$\frac{7}{2}_{2}^{+}$  &     6.6       &    0.35       &             \\

$\frac{3}{2}_{2}^{+}$ &$\frac{1}{2}_{1}^{+}$  &     8.5       &     2.5       &     2.5(5)  \\
$\frac{3}{2}_{2}^{+}$ &$\frac{3}{2}_{1}^{+}$  &  $<$0.01      &     3.9       &$23^{+5}_{-7}$\\
$\frac{3}{2}_{2}^{+}$ &$\frac{7}{2}_{1}^{+}$  &     8.1       &    11.9       &             \\

$\frac{7}{2}_{2}^{+}$ &$\frac{5}{2}_{1}^{+}$  &     24.1      &    10.1       &             \\
$\frac{7}{2}_{2}^{+}$ &$\frac{3}{2}_{1}^{+}$  &     7.9       &     3.5       &             \\
$\frac{7}{2}_{2}^{+}$ &$\frac{7}{2}_{1}^{+}$  &     0.35      &     2.0       &             \\
$\frac{7}{2}_{2}^{+}$ &$\frac{5}{2}_{2}^{+}$  &     9.0       &     7.5       &             \\
$\frac{7}{2}_{2}^{+}$ &$\frac{3}{2}_{2}^{+}$  &     1.7       &     1.1       &             \\

  \end{tabular}
  \label{SMBE2Table}
  \footnotetext[1]{Transition strength from Ref.~\cite{McGowan1958} evaluated with RULER~\cite{Burrows1984}.}
  \footnotetext[2]{Transition strength calculated from the measured mean lifetime and branching ratios found in this work. The Nuclear Data Sheet value is 4.4(8)~W.u~\cite{Blachot2009} for the ground state transition, 5.9(15)~W.u. with the new branching ratios.}
\end{table}
\end{ruledtabular}

More important for the evaluation of the emergence of collectivity in $^{111}$Cd are the $E2$ transition rates shown in Table~\ref{SMBE2Table}. Generally, the stronger $E2$ transitions in the range of 10 to 20~W.u. are strong in both theory and experiment. It is also noteworthy that the predicted $E2$ strength has the correct magnitude - no attempt has been made to tune the effective charges to better describe experiment.

The strongly Coulomb-excited states have $\sim10$~W.u. transitions to the ground state; these are shown in bold in Table~\ref{SMBE2Table}. The $E2$ decay of the $5/2_{1}^{+}$ isomeric state is well described by the theory. However, the $5/2^{+}_{2}\rightarrow 1/2^{+}_{\rm{g.s.}}$ is underpredicted and the $5/2^{+}_{3}\rightarrow 1/2^{+}_{\rm{g.s.}}$ decay is overpredicted. This suggests that the mixing between these $5/2^{+}$ states may not be fully captured in the calculations. Even so, the shell model correctly predicts an isomeric $5/2^{+}_{1}$ state and then two additional $5/2^{+}$ with strong $E2$ coupling to the ground state.

Two transitions in Table~\ref{SMBE2Table} are predicted to be very weak ($<1$~W.u.) but have experimental values of the order of $20$~W.u.: in the case of the $3/2^{+}_{1}\rightarrow5/2^{+}_{1}$ transition, the shell model predicts an almost pure $M1$ transition, consistent with an observed mixing ratio of $\delta \sim 0.1$, and the experimental uncertainty on the $E2$ strength is so large that there is no compelling discrepancy between theory and experiment. The case of the $3/2^{+}_{2}\rightarrow3/2^{+}_{1}$ transition depends on uncertain branching ratios and a $B(E2)$ value reported only in a conference proceedings~\citep{Andreev1975}.

Additional and more accurate experimental data on transition rates are clearly needed to enable a more detailed and critical evaluation of the developing collectivity in $^{111}$Cd.

The general qualitative agreement between the shell model and the particle-rotor model for $g$~factors in this weakly collective nucleus can be attributed to Lawson's observation~\cite{Lawson1980} that Nilsson wavefunctions can provide an excellent approximation to those emerging from a Hartree-Fock calculation. In cases where there are only a few particles outside the closed shell, the Nilsson potential approximates the independent motion of nucleons in an oscillator potential under the influence of an attractive quadrupole-quadrupole effective interaction. This scenario also describes the underlying quadrupole-collectivity-driving features of the shell-model calculation.


\section{Conclusion} \label{ConclusionSection}

The structure of $^{111}$Cd has been studied through angular correlation, lifetime, and $g$-factor measurements. The $5/2^{+}$ spin-parity assignment of the 753-keV state was confirmed. It is clearly not the analogue of the $3/2^{+}$ \mbox{681-keV} state in $^{113}$Cd. It is also suggested that the reported $3/2^{+}$ state at 755~keV~\cite{Blachot2009} is a misidentification of the observed $5/2_{3}^{+}$ state. Non-observation of a $3/2^{+}$ state near 700~keV in $^{111}$Cd is difficult to explain within the particle-vibration scheme. Further, the nature of the observed $5/2^{+}_{3}$ state has been shown to be inconsistent with the predictions of the particle-vibration model. In contrast, the observed states do match those expected within a particle-rotor description.

A simple particle-rotor framework provides a natural explanation of both the electric quadrupole moments and magnetic properties of the observed states which cannot be readily explained in the particle-vibration model. However, the particle-rotor approach is not without its own limitations. For example, the energy spectrum cannot be described by rigid rotations. More rigorous consideration of the stiffness of deformation in cadmium nuclei is needed. Further theoretical investigation of weakly-deformed nuclei from diverse theoretical perspectives is necessary for a more complete understanding.

Large-scale shell-model calculations have been performed for $^{111}$Cd and it has been shown that the signs and magnitudes of the experimental $g$~factors are reproduced using standard single-particle $g$~factors. The low-lying level scheme also agrees well with experiment. These calculations also produce transition strengths in overall agreement with the observed $E2$ transition strengths for the primary low-excitation transitions. It is suggested, following Lawson~\cite{Lawson1980}, that the agreement between the particle-rotor description and the shell model for the $g$~factors stems from the fact that the Nilsson wavefunctions at small deformation provide a good approximation to the shell model. Coulomb excitation studies to measure more extensive electromagnetic matrix elements are needed for a more critical comparison of theory and experiment. Overall, the results of the shell-model calculations are promising and represent a new opportunity to gain further insight into these weakly collective nuclei from the microscopic perspective.


\begin{acknowledgments}

 The authors are grateful to the academic and technical staff of the Department of Nuclear Physics (Australian National University) and the Heavy Ion Accelerator Facility for their support. This research was supported in part by the Australian Research Council grant numbers DP120101417, DP130104176, DP140102986, DP140103317,
DP170101673, and LE150100064. B.J.C., A.A.,
J.T.H.D., M.S.M.G., and T.J.G. acknowledge support of the Australian
Government Research Training Program. Support for the
ANU Heavy Ion Accelerator Facility operations through
the Australian National Collaborative Research Infrastructure
Strategy (NCRIS) program is acknowledged.

\end{acknowledgments}

\bibliography{g_factors_in_111Cd_Bibliography}

\end{document}